\documentclass[%
 reprint,
superscriptaddress,
 amsmath,amssymb,
 aps,
prl,
]{revtex4-2}

\usepackage{dcolumn}
\usepackage{chemformula}
\usepackage[T1]{fontenc}
\usepackage[utf8]{inputenc}
\usepackage{bbold}
\usepackage{afterpage}
\usepackage{float}
\usepackage{amsmath}
\usepackage{amssymb}
\usepackage{graphicx}
\usepackage[unicode=true,pdfusetitle,
bookmarks=true,bookmarksnumbered=false,bookmarksopen=false,
breaklinks=false,pdfborder={0 0 1},backref=false,colorlinks=false]
{hyperref}
\usepackage{xcolor}
\usepackage[title]{appendix}
\usepackage{physics}
\usepackage{soul}
\usepackage{dsfont}
\usepackage{orcidlink}
\usepackage[normalem]{ulem}
\usepackage{siunitx}
\usepackage{multirow}
\usepackage{capt-of}
\usepackage{tabularx}
\hypersetup{
    colorlinks,
    linkcolor={red!50!black},
    citecolor={blue!50!black},
    urlcolor={blue!80!black}
}

\newcommand{\markup}[1]{\textcolor{black}{#1}}
\newcommand{\mymarkup}[1]{\textcolor{black}{#1}}

\begin{document}


\newcommand{\tmd}{\gls{tmd}}
\newcommand{\dft}{\gls{dft}}
\newcommand{\mbpt}{\gls{mbpt}}
\newcommand{\rpa}{\gls{rpa}}
\newcommand{\bz}{\gls{bz}}
\newcommand{\lcao}{\gls{lcao}}
\newcommand{\rhs}{\gls{rhs}}
\newcommand{\lhs}{\gls{lhs}}
\newcommand{\hf}{\gls{hf}}
\newcommand{\hbn}{\gls{hbn}}
\newcommand{\bse}{\gls{bse}}
\newcommand{\gs}{\gls{gs}}
\newcommand{\soc}{\gls{soc}}

\newcommand{\abinit}{\emph{ab~initio}~}
\newcommand{\cti}{\textbf{continuous} translation invariance}
\newcommand{\dti}{\textbf{discrete} translation invariance}
\newcommand{\qp}{quasi-particle}
\newcommand{\wf}{wave-function}





\newcommand{\Ha}{\hat{H}}
%
\newcommand{\Ho}{\hat{H}_0}
%




\newcommand{\bzero}{\vb{0}}

\newcommand*{\rttensor}[1]{\overline{\overline{#1}}}

\newcommand{\bR}{\vb{R}}

\newcommand{\bQ}{\vb{Q}}

\newcommand{\bt}{\vb{t}}

\newcommand{\bb}{\vb{b}}

\newcommand{\bG}{\vb{G}}

\newcommand{\bn}{\vb{n}}

\newcommand{\br}{\vb{r}}

\newcommand{\bx}{\vb{x}}

\newcommand{\by}{\vb{y}}

\newcommand{\bC}{\mathbf{C}}

\newcommand{\bv}{\vb{v}}


\newif\ifusebm
\usebmtrue 

\ifusebm
    \providecommand{\bE}{\bm{E}}

    \providecommand{\bA}{\bm{A}}
    \providecommand{\hA}{\hat{\bm{A}}}
    \providecommand{\nhA}{\bm{A}} 

    \providecommand{\bB}{\bm{B}}

    \providecommand{\bD}{\bm{D}}

    \providecommand{\bA}{\bm{A}}
    \providecommand{\hA}{\hat{\bm{A}}}
    \providecommand{\nhA}{\bm{A}} 

    \providecommand{\bB}{\bm{B}}

    \providecommand{\bH}{\bm{H}}

    \providecommand{\bJ}{\bm{J}}

    \providecommand{\bj}{\bm{j}}

    \providecommand{\bP}{\bm{P}}

    \providecommand{\bM}{\bm{M}}

\else
    \providecommand{\bE}{\vb{E}}

    \providecommand{\bD}{\vb{D}}

    \providecommand{\bA}{\vb{A}}
    \providecommand{\hA}{\hat{\vb{A}}}
    \providecommand{\nhA}{\vb{A}} 

    \providecommand{\bB}{\vb{B}}

    \providecommand{\bH}{\vb{H}}

    \providecommand{\bJ}{\vb{J}}

    \providecommand{\bj}{\vb{j}}

    \providecommand{\bP}{\vb{P}}

    \providecommand{\bM}{\vb{M}}
\fi

\newcommand{\bk}{\vb{k}}
\newcommand{\bq}{\vb{q}}
\newcommand{\bp}{\vb{p}}

\newcommand{\bK}{\vb{K}}

\newcommand{\bex}{\vb{e}_x}
\newcommand{\bey}{\vb{e}_y}
\newcommand{\bez}{\vb{e}_z}

\newcommand{\bux}{\vb{u}_x}
\newcommand{\buy}{\vb{u}_y}
\newcommand{\buz}{\vb{u}_z}

\newcommand{\bhe}{\hat{\vb{e}}}

\newcommand{\be}{\vb{e}}

\newcommand{\ii}{\mathrm{i}}

\newcommand{\ee}{\mathrm{e}}

\newcommand{\ex}[1]{\mathrm{e}^{#1}}

\newcommand{\ei}[2][]{\ee^{#1 \ii #2}} 

\newcommand{\TE}{\mathrm{TE}}
\newcommand{\TM}{\mathrm{TM}}

\newcommand{\adagger}{a^{\dagger}}
\newcommand{\bdagger}{b^{\dagger}}
\newcommand{\cdagger}{c^{\dagger}}

\newcommand{\hw}{\hbar \omega}


\newcommand{\dt}{\dd{t}}

\newcommand{\dr}[1][]{\dd[#1]{\br}}

\newcommand{\dx}[1][]{\dd[#1]{\bx}}
\newcommand{\dy}[1][]{\dd[#1]{\by}}

\newcommand{\dk}[1][]{\dd[#1]{\bk}}
\newcommand{\dpp}[1][]{\dd[#1]{\bp}}
\newcommand{\dq}[1][]{\dd[#1]{\bq}}

\newcommand{\lap}{\laplacian} 
\newcommand{\rot}{\bm{\nabla} \times} 


\newcommand{\pdt}[1][]{\pdv{#1}{t}} 
\newcommand{\pdti}[1][]{\pdv*{t}} 

\newcommand{\pdx}[1][]{\pdv{#1}{x}} 
\newcommand{\pdy}[1][]{\pdv{#1}{y}} 
\newcommand{\pdz}[1][]{\pdv{#1}{z}} 

\newcommand{\delt}{\partial_t} 
\newcommand{\delx}{\partial_x} 
\newcommand{\dely}{\partial_y} 
\newcommand{\delz}{\partial_z} 

\newcommand{\pdd}[2][]{\pdv[2]{#1}{#2}} 

\newcommand{\pddt}[1][]{\pdd[#1]{t}} 

\newcommand{\pddx}[1][]{\pdd[#1]{x}} 
\newcommand{\pddy}[1][]{\pdd[#1]{y}} 
\newcommand{\pddz}[1][]{\pdd[#1]{z}} 

\newcommand{\ddelt}{\partial^2_t} 
\newcommand{\ddelx}{\partial^2_x} 
\newcommand{\ddely}{\partial^2_y} 
\newcommand{\ddelz}{\partial^2_z} 

\newcommand{\tdt}[1][]{\dv{#1}{t}} 


\newcommand{\Kd}[1]{\delta_{#1}} 
\newcommand{\Kdc}[2]{\delta_{#1,#2}} 

\newcommand{\chirpa}{\chi^{\mathrm{\scriptscriptstyle RPA}}}
\newcommand{\chiirr}{\chi^{\mathrm{\scriptscriptstyle irr}}}


\renewcommand{\wp}{\omega_{\text{p}}} 

\newcommand{\EF}{E_F} 
\renewcommand{\vb}[1]{\mathbf{#1}}

\newcommand{\MoS}{\ch{MoS2}}
\providecommand{\pnl}[1]{{\textcolor{black}{\mbox{\textbf{#1}}}}}


\newcommand{\peni}[1]{\textsuperscript{\textcolor{red!70!black}{peni}}  \textcolor{red!70!black}{[#1]}
}

\newcommand{\ct}[1]{\textsuperscript{\textcolor{green!70!black}{ct}}  \textcolor{green!70!black}{[#1]}
}

\newcommand{\namo}[1]{\textsuperscript{\textcolor{red}{namo}}  \textcolor{red}{[#1]}
}

\newcommand{\alex}[1]{\textsuperscript{\textcolor{blue!70!black}{alex}}  \textcolor{blue!70!black}{[#1]}
}
\newcommand{\juanjo}[1]{\textsuperscript{\textcolor{orange!70!black}{juanjo}}  \textcolor{orange!70!black}{[#1]}
}

\title{Microscopic screening theory for excitons in two-dimensional materials: A bridge between effective models and \textit{ab initio} descriptions}
\author{Pedro Ninhos\,\orcidlink{0000-0002-1143-7457}}
\affiliation{POLIMA---Center for Polariton-driven Light--Matter Interactions, University of Southern Denmark, Campusvej 55, DK-5230 Odense M, Denmark}
\affiliation{Department of Condensed Matter Physics, Universidad Aut\'{o}noma de Madrid, 28049 Madrid, Spain}

\author{Alejandro J. Ur\'{i}a-\'{A}lvarez\,\orcidlink{0000-0001-6668-7333}}
\affiliation{Department of Condensed Matter Physics, Universidad Aut\'{o}noma de Madrid, 28049 Madrid, Spain}
\affiliation{Condensed Matter Physics Center (IFIMAC) and Instituto Nicol\'{a}s Cabrera (INC), Universidad Aut\'{o}noma de Madrid, 28049 Madrid, Spain}
\affiliation{Departamento de F\'isica, Universidad de Oviedo, 33007 Oviedo, Spain}

\author{Christos~Tserkezis\,\orcidlink{0000-0002-2075-9036}}
\affiliation{POLIMA---Center for Polariton-driven Light--Matter Interactions, University of Southern Denmark, Campusvej 55, DK-5230 Odense M, Denmark}
\affiliation{Danish Institute for Advanced Study, University of Southern Denmark, Campusvej 55, DK-5230 Odense M, Denmark\\
\textup{*Corresponding author: juanjose.palacios@uam.es}}

\author{N. Asger Mortensen\,\orcidlink{0000-0001-7936-6264}}
\affiliation{POLIMA---Center for Polariton-driven Light--Matter Interactions, University of Southern Denmark, Campusvej 55, DK-5230 Odense M, Denmark}
\affiliation{Danish Institute for Advanced Study, University of Southern Denmark, Campusvej 55, DK-5230 Odense M, Denmark\\
\textup{*Corresponding author: juanjose.palacios@uam.es}}

\author{Juan Jos\'{e} Palacios$^*$\,\orcidlink{0000-0003-2378-0866}}
\affiliation{Department of Condensed Matter Physics, Universidad Aut\'{o}noma de Madrid, 28049 Madrid, Spain}
\affiliation{Condensed Matter Physics Center (IFIMAC) and Instituto Nicol\'{a}s Cabrera (INC), Universidad Aut\'{o}noma de Madrid, 28049 Madrid, Spain}

\date{March 2025}
\begin{abstract}
\hspace{5.5cm}\textbf{ABSTRACT}\\
We present a computational approach for exciton calculations in two-dimensional (2D) materials within the Bethe--Salpeter equation (BSE) framework, employing an atomistic description with point-like orbitals. Unlike widespread efficient calculations that rely on classical or effective interaction models, such as the Rytova--Keldysh model, our method incorporates quantum screened interactions. By explicitly computing the 2D dielectric function at the random-phase approximation level, we capture screening effects beyond such approximations with an accuracy akin to first-principles methods. Consequently, we can realistically estimate excitonic binding energies with a bearable computational cost. A detailed account of the various convergence parameters sheds light on a possible cause of the large dispersion of binding energies reported in the literature using first-principles $GW$/BSE implementations. This work thus provides an alternative pathway towards efficient and faithful dielectric screening and exciton computations in low-dimensional materials.
\end{abstract}

\maketitle

\section*{Introduction}

Within a quantum mechanical framework, the optoelectronic properties of solid-state materials can generally be described in terms of electronic excitations from occupied states to higher-energy unoccupied states~\cite{Fox_Fox_2010,Kittel_2004}. In metals or in materials without a bandgap, these transitions between single-particle states are typically intraband in nature when the excitation energies are sufficiently low. 
In contrast, semiconducting or insulating materials in their ground state require energies typically greater than the single-particle electronic bandgap.
Interestingly, the optical response of semiconductors and insulators is non-trivial for frequencies within the range of energies spanning such electronic bandgap~\cite{Koch_2009} where single-particle states are absent. Yet, the system can host quasi-particles known as excitons---bound electron--hole pairs that form at sub-gap energies. This bound state is formally a many-body state; it can only be \mymarkup{predicted} when electron--electron interactions are \mymarkup{included in the many-body static Hamiltonian} beyond the single-particle or mean-field level~\cite{Mahan,Cohen_Louie_2016}.

Excitons play a central role in determining the optical properties of low-dimensional crystals, such as hexagonal boron nitride (hBN)~\cite{Ferreira_Chaves_Peres_Ribeiro_2019,Cassabois_2019} or transition metal dichalcogenide (TMD) materials~\cite{PhysRevB.86.115409,Berkelbach_Hybertsen_Reichman_2013}. The current interest in excitons within two-dimensional (2D) materials is partly driven by the experimental demonstration of exciton-based light–matter interactions at room temperature~\cite{Flatten_He_Coles_Trichet_Powell_Taylor_Warner_Smith_2016,Morozov_Wolff_Mortensen_2021,Fang_Yao_Wang_2023}. Unlike in their three-dimensional (3D) counterpart, excitons in 2D materials are long-lived and have considerable binding energies ranging from a fraction of electron-volt (eV) to a couple of eV.
The large exciton binding energies in 2D materials, compared with those in 3D systems, can be explained by the electrostatic screening being less effective in 2D than in 3D~\cite{Chernikov_2014}. Due to the reduced dimensionality, the dielectric response of the material essentially affects the electric field lines within it, whereas those off-plane are less affected by the material response. As a consequence, the relative permittivity of a 2D material cannot be described by a single permittivity constant, which poses a challenge from a theoretical standpoint.
Somehow, the theory must incorporate the fact that the relative permittivity varies in space and promote it from a constant to a function---the dielectric function. It plays a central role in describing how electric charges inside a material respond to external electric field perturbations within the linear response.

The strong dependence of the dielectric function on space is what confers so-called nonlocal effects to the screening material properties~\cite{Cudazzo_Tokatly_Rubio_2011}. In other words, the dielectric function in reciprocal space exhibits  dispersion with strong dependence on the wavevector. To model this dependence, the analytical Rytova--Keldysh model~\cite{keldysh2024coulomb} has been put forward solely on classical grounds, providing an intuitive interpretation for the screened Coulomb potential felt between two electric charges in a 2D crystal. Due to its simple expression in reciprocal space, which follows a linear dependence on the momentum, it offers as well an analytical expression for the Coulomb potential in real space and is widely used in exciton calculations  combining analytical with numerical methods~\cite{Palacios.PhysRevB.91.245421,Gomes_Trallero-Giner_Vasilevskiy_2021}. Nonetheless, it has some shortcomings. In the light of classical physics, even though one can predict the behavior of the dielectric function in the low-momentum limit, it is not possible to estimate the model parameter appearing in its formula without the help from quantum mechanics. Moreover, it only captures the screening accurately in the long-wavelength limit. To go beyond that, one has to resort to first-principles methods to obtain the full dependence on momentum. Also called \textit{ab initio} methods, they permit a description of screening and excitons within a single framework, albeit completely computational.
This framework relies first on knowledge of the ground state of the system.
To date, the most powerful framework for this purpose is density functional theory (DFT)~\cite{Hohenberg_Kohn_1964,Kohn_Sham_1965}, 
a theory that enables, in principle, the determination of ground state properties of both finite and extended systems based only on the atomic structure under consideration.

Through DFT calculations, we can obtain an approximate single-particle description of the material, on top of which we can do post-processing calculations to obtain the excited states. As DFT usually underestimates the bandgap; before calculating the exciton, this needs to be corrected to bring it closer to the experimental value. The tools of many-body perturbation theory (MBPT)~\cite{Mahan,Bruus_Flensberg_2004} provide a possible way to correct the quasiparticle energies of our many-electron system. The $GW$~\cite{Hybertsen_1986,Martin_Reining_Ceperley_2016} approximation is one of them, which accounts for dynamical screening and can accurately correct the electronic bandgap. Together with the Bethe--Salpeter equation (BSE) within MBPT, the $GW$+BSE~\cite{Louie_2000,Martin_Reining_Ceperley_2016,Diana_PhD_thesis,da_Jornada_PhD_thesis} method can in one sitting predict the quasiparticle energies and the excitonic levels with great accuracy~\cite{Martin_Reining_Ceperley_2016,Louie_2000}. Unfortunately, such methods are deemed computationally too heavy. Often, it suffices to extract the numerical macroscopic dielectric function from the DFT calculations, and then fit it to a linear model to recover the Rytova--Keldysh dielectric function. Then,  exciton calculations follow, typically by using the BSE on top of a tight-binding description of the material~\cite{Ridolfi_2018,Galvani_2016,Alex_Xatu_2024,Henriques_Epstein_Peres_2022,Quintela_Henriques_Tenorio_Peres_2022} or through the effective Wannier equation~\cite{Wannier_1937,Blount_1962,Ventura_2019}. Nevertheless, despite the extensive literature, there is a methodology gap that the present work aims to fill. 

In this manuscript, we explore a middle ground in-between analytical models and \textit{ab initio}~descriptions of screening and excitons in 2D extended periodic systems. Taking the reduced dimensionality of the 2D excitonic problem to our advantage, and without compromising a microscopically detailed description of screening, we implement an efficient atomistic description of the microscopic dielectric function based on a strictly 2D formalism. 
The theory developed herein allows for the evaluation of the full static dielectric response at a much lower computational cost than first-principles methods, capturing both long- and short-wavelength limits (the former being where effective models usually operate). This is done with no apparent sacrifice of accuracy, since our results can match those reported in the \textit{ab initio} literature.
Furthermore, the extension of our strict 2D description to a quasi-two-dimensional (Q2D) theory allows for: (i)~justifying the zero-thickness approximation done in the strict 2D formalism and investigating its regimes of validity; (ii) improving the dielectric response encoded in the effective macroscopic dielectric function, matching the \textit{ab initio}~results of the literature with excellent qualitative and quantitative agreement. All of this, still at a significantly smaller fraction of the computational cost of first-principles methods.

To benchmark and validate our results, we apply the theory and the implementation to two paradigmatic materials, hexagonal boron nitride (\ch{hBN}) and molybdenum disulfide (\ch{MoS2}). For both materials, we obtain an effective 2D dielectric function rather close to the \textit{ab initio}~one, recovering  a quantitatively accurate Rytova--Keldysh linear behavior in the long-wavelength limit. As a direct application of our numerically obtained dielectric function (and matrix), 
we calculate excitonic states through the BSE as implemented in the XATU code~\cite{Alex_Xatu_2024}.
Not only are the results for the exciton energies herein determined compatible with those obtained using first-principles methods,
the efficiency of our methodology also allows for a detailed convergence analysis of these energies with the different calculation parameters. Such a complete convergence analysis, including the local-field effects in the interaction kernel of the BSE, is not always feasible, particularly when using \textit{ab initio}~calculations due to prohibitive computational costs, and deserves due attention.

\section*{Results}
\setcounter{subsection}{0}
\subsection{Band structure}
\label{subsec:Band_structure_results}

\begin{figure}[h]
    \centering
    \includegraphics[width=1.0\linewidth]{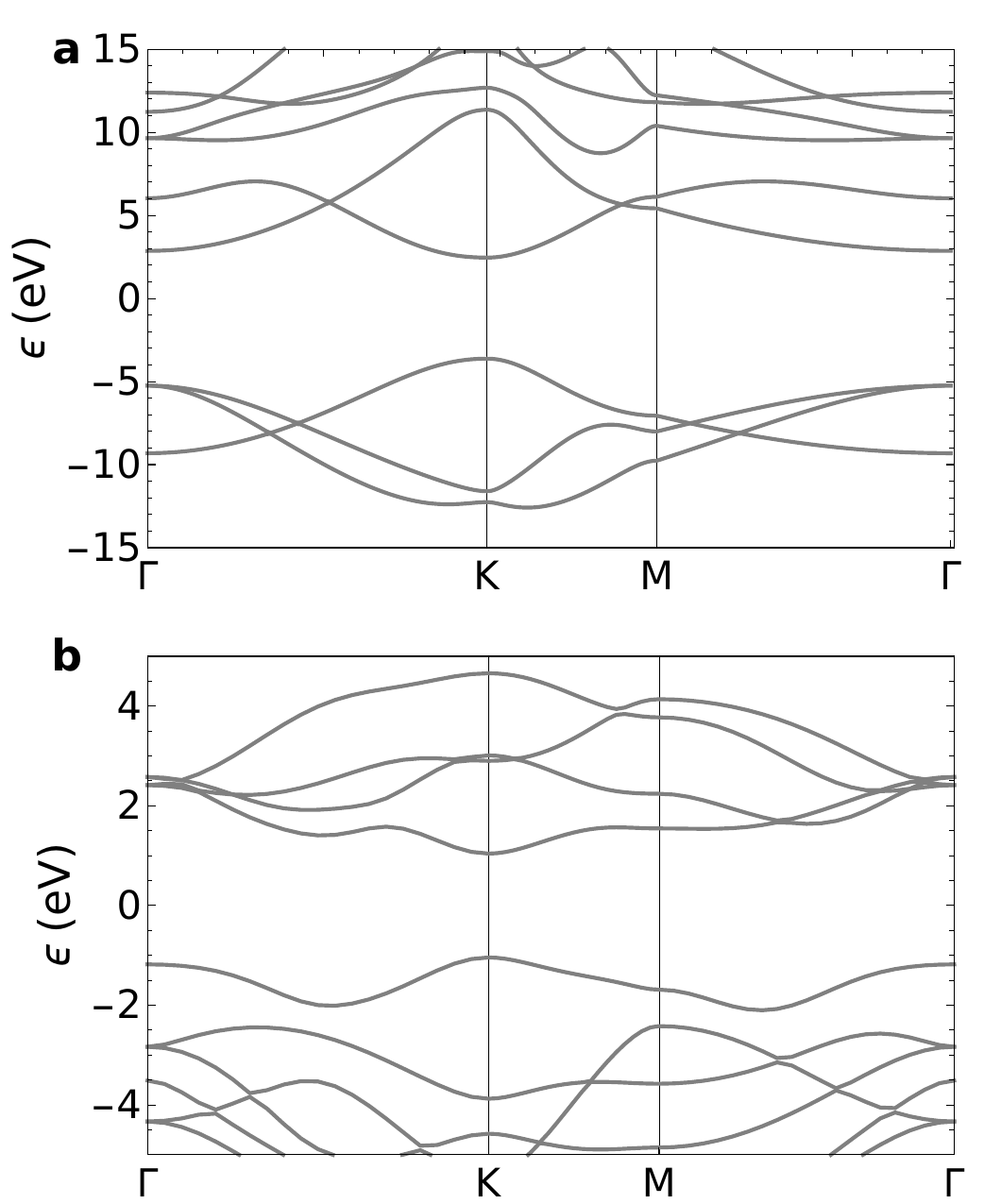}
    \caption[justification=justified]{\textbf{Band structures of monolayer \ch{hBN} and monolayer \ch{MoS2}.} Band structure of monolayer \ch{hBN} in panel \pnl{a} and of monolayer\ch{MoS2} in panel \pnl{b}, obtained through DFT calculations using the exchange--correlation functional HSE06. For both materials, the bandgap is located at the K point of the BZ, and it is $\Delta=6.08$\,eV for hBN and $\Delta=2.08$\,eV for \ch{MoS2}. For a better visualization, only the first few bands around the bandgap of each material are displayed.}
    \label{fig:hBN_&_MoS2_bands}
\end{figure}

Excited states of materials are linked to their ground state properties. In an infinite periodic structure, or crystal for short, the single-particle electronic states or Bloch states are typically described by a band index $n$ and a momentum vector $\mathbf{k}$. The set of all quasiparticle electronic energies $\epsilon_{n \mathbf{k}}$ for every possible $n$ and $\mathbf{k}$ constitute the band structure of the material. Even if computed through DFT, the Bloch states $|n \mathbf{k} \rangle$ and energies $\epsilon_{n \mathbf{k}}$ commonly constitute a basis for approaching the truly interacting many-body problem. We present first the electronic band structure for two 2D gapped materials, \ch{hBN} as an insulator and \ch{MoS2} as a semiconductor. Figure~\ref{fig:hBN_&_MoS2_bands} displays the band structure of \ch{hBN} and \ch{MoS2}~in panels~\pnl{a} and \pnl{b}, respectively. The band structure of \ch{hBN} was obtained with the commercial package CRYSTAL~\cite{CRYSTAL_2023} through DFT calculations with a $48\times 48\times 1$ $\mathbf{k}$-grid within a framework of localized functions using a Gaussian basis with 69 orbitals (69 bands in total, of which 6 are valence bands and 63 are conduction bands. The band structure features a direct bandgap of $\Delta = 6.08$\,eV at the high-symmetry point $\mathrm{K}$ in the Brillouin zone (BZ). For the exchange--correlation functional, the hybrid functional HSE06~\cite{Heyd_Scuseria_Ernzerhof_2003} was used, which allows for a much better estimation of the electronic bandgap than using, for instance, the local-density approximation (LDA).  The lattice constant is $a=2.51$\,{\AA} and  the inter-layer distance used was $L_{\perp} = 500$\,{\AA}. CRYSTAL also provides the Fock matrices, from which we perform our own computation of $|n \mathbf{k} \rangle$. As explained in the Methods~\ref{subsec:Chi&epsilon_Methods} section, knowledge of the eigenenergies and eigenfunctions allows the calculation of the dielectric function.

The band structure of \ch{MoS2} was obtained similarly, without including spin--orbit coupling (SOC), since we have
verified that SOC is irrelevant to compute its screening properties. More specifically, the difference in the matrix elements of the polarizability is seen only in the sixth significant decimal place. 
The lattice constant for \ch{MoS2}~is $a=3.16$\,{\AA} with the \ch{Mo} atoms in-plane, and the \ch{S} atoms with an off-plane displacement of $1.585$\,{\AA}.
The inter-layer distance used was $L_{\perp} = 500$\,{\AA}, and the basis has a size of 62 atomic orbitals, meaning that the Fock matrices are $62$ by $62$. Therefore, this model has in total 62 bands, of which 23 are valence bands and 39 are conduction bands.
This material has the same hexagonal crystalline structure as \ch{hBN} even though it has three atoms in the unit cell instead of two, and its bandgap of $\Delta = 2.08$\,eV is as well located at the high-symmetry point K in the BZ.

In general, hybrid functionals allow for a very good estimate of the bandgap with lower computational effort than $GW$ calculations. Nonetheless, as we will observe in the Results~\ref{subsec:dielectric_function_results} section, the calculation of the dielectric function is not grandly affected by the exchange--correlation functional used in the DFT calculations, as our results agree very well with other models in the literature for the same materials with the band structure obtained within LDA.

\subsection{Polarizability}
\label{subsec:Results_Polarizability}

Before computing the dielectric function matrix elements, one has to ensure that the irreducible non-interacting polarizability that we compute through
\begin{equation}
\begin{split}
\label{eq:Chi_0_static_results}
    &\chi^0_{\mathbf{G} \mathbf{G}'} (\mathbf{q}) = \frac{2}{N} \sum_{vc,\mathbf{k}\sigma} \frac{I^{\mathbf{G}}_{c \mathbf{k} , v \mathbf{k} + \mathbf{q}} \left(I^{\mathbf{G}'}_{c \mathbf{k} , v \mathbf{k} + \mathbf{q}} \right)^*}{\epsilon_{v \mathbf{k}+\mathbf{q}}-\epsilon_{c \mathbf{k}}}
\end{split}
\end{equation}
\noindent is well converged for any generic momentum $\mathbf{q}$ in the BZ.
Here, $\sigma$ accounts for spin, the numerator is a product of matrix elements of the form $I^{\mathbf{G}}_{n\mathbf{k}, n' \mathbf{k}'} \equiv \langle n,\mathbf{k} | \mathrm{e}^{-\mathrm{i}(\mathbf{k}' - \mathbf{k} + \mathbf{G}) \cdot \mathbf{r}} | n',\mathbf{k}' \rangle$, $\mathbf{G}$ and $\mathbf{G}'$ are generic vectors of the reciprocal lattice, $v/c$ denotes valence/conduction band, and $\epsilon_{n \mathbf{k}}$ represents the quasiparticle energies as usual. The sum in momentum $\mathbf{k}$ is over the whole BZ. Details on the numerical evaluation of this expression are presented in the Methods~\ref{subsec:Chi&epsilon_Methods} section.
A key approximation in this work is the point-like orbital approximation, where we ignore altogether the spatial details of the atomic orbitals; this allows us to work out an analytical expression for the matrix elements, which can be evaluated numerically without the need of using any additional library to perform fast Fourier transforms (FFTs), and with negligible computational effort.

\begin{figure}[!htb]
    \centering
    \includegraphics[width=1.0\linewidth]{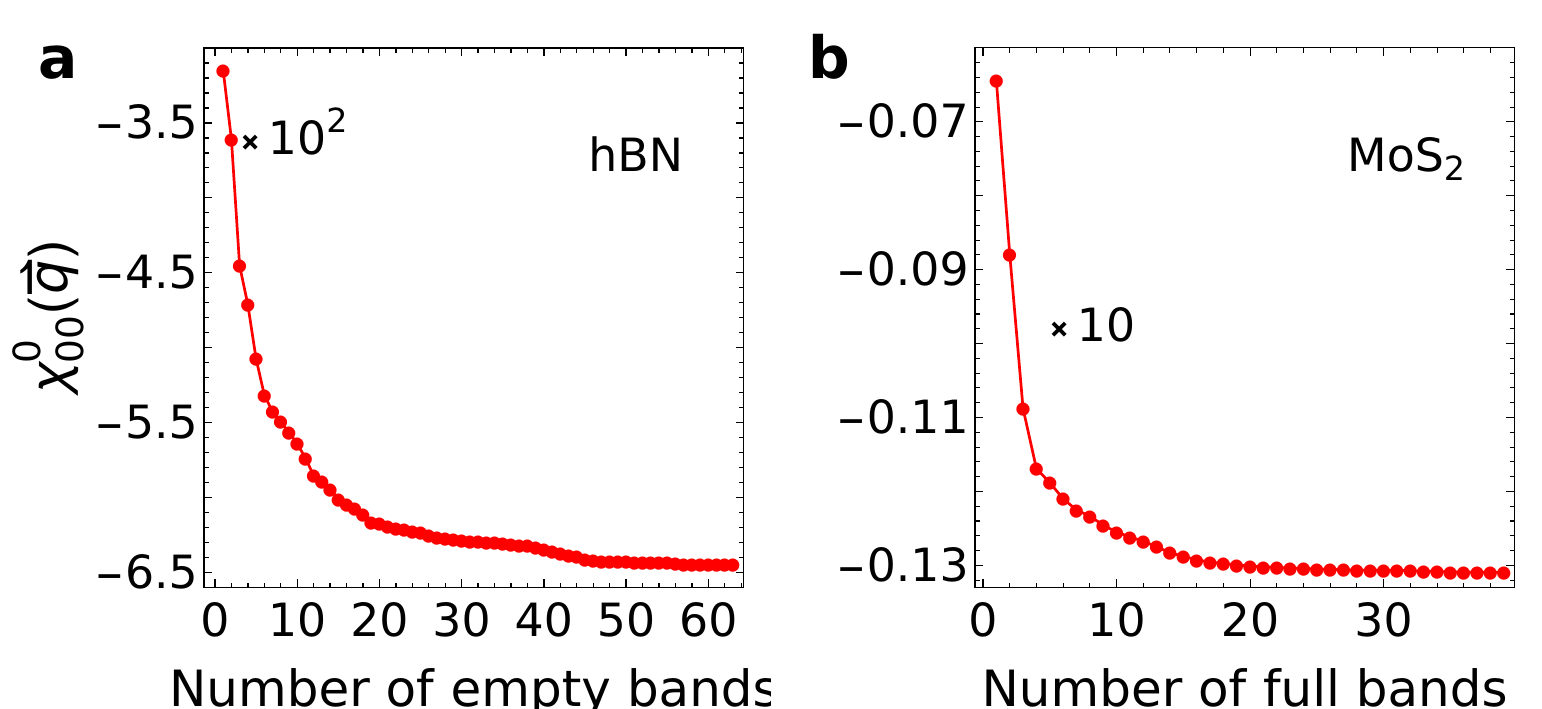}
    \caption[justification=justified]{\textbf{Head matrix element of the polarizability.} Convergence of the head matrix element of the non-interacting polarizability with the number of conduction bands included in the sum in Eq.~\eqref{eq:Chi_0_static_results}. Panel \pnl{a} is for \ch{hBN} and $\mathbf{q} = (0.5,0)$\,{\AA}$^{-1}$, while panel \pnl{b} is for \ch{MoS2}~and $\mathbf{q} = (0.2,0)$\,{\AA}$^{-1}$. Both polarizability elements are converged with the number of momentum points in the BZ and appear in eV$^{-1}$.}
    \label{fig:Chi0_convergence}
\end{figure}

Although the convergence with the number of points $N$ in the BZ is not recorded in this manuscript, we have verified that a sparse mesh is sufficient, where only around $N \sim 17^2$ momentum points ($\sqrt{N}$ in each direction) are needed, without much difference from material to material.
We have also verified the convergence with the number of conduction bands, and the results are displayed in Fig.~\ref{fig:Chi0_convergence}. We test the convergence of the $\mathbf{G} = \mathbf{G}' = \mathbf{0}$ element of the non-interacting polarizability $\chi^0_{\mathbf{0}\mathbf{0}}(\mathbf{q})$, usually called head element in the literature, as per Eq.~\eqref{eq:Chi_0_static_results} with the number of conduction bands included in that sum. Figures~\ref{fig:Chi0_convergence}\pnl{a} and \ref{fig:Chi0_convergence}\pnl{b} correspond to \ch{hBN} and \ch{MoS2}, respectively, for a specific $\mathbf{q} \in \mathrm{BZ}$. 
We also tested the convergence of other elements of the non-interacting polarizability matrix (not shown), off-diagonal ones for instance, by checking the convergence of both real and imaginary parts separately (whenever the imaginary part is non-zero).

\begin{figure}[!htb]
    \centering
    \includegraphics[width=1.0\linewidth]{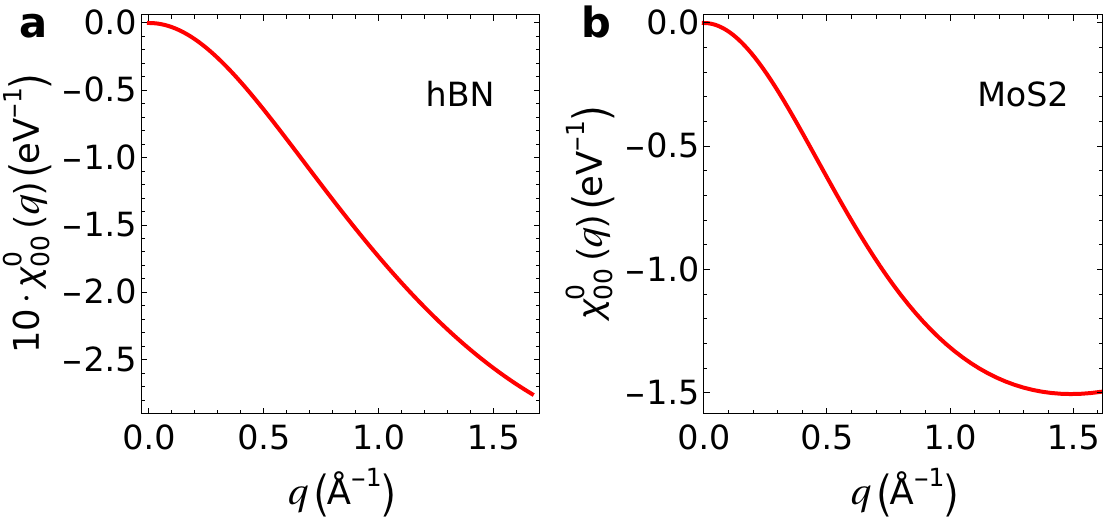}
    \caption[justification=justified]{\textbf{Head matrix element of the polarizability as a function of momentum.} Head matrix element of the irreducible polarizability $\chi^0_{\mathbf{0} \mathbf{0}}(\mathbf{q})$, as per Eq.~\eqref{eq:Chi_0_static_results}, as a function of momentum magnitude, for \ch{hBN} in panel \pnl{a} (scaled by a factor of $10$) and for \ch{MoS2}~in panel \pnl{b}. The polarizability was computed in both cases along the symmetry line $\Gamma - \mathrm{K}$. The momentum $q=|\mathbf{q}|$ is in {\AA}$^{-1}$ and the polarizability in eV$^{-1}$. Note the different scales for the vertical and horizontal axes in each plot. In both panels, all the valence and conduction bands were included in the sum \eqref{eq:Chi_0_static_results}.}
    \label{fig:Chi0_vs_q_hBN_&_MoS2}
\end{figure}

To further validate the calculated polarizability, in Fig.~\ref{fig:Chi0_vs_q_hBN_&_MoS2} we study the behavior of the $\chi^0_{\mathbf{0}\mathbf{0}}(\mathbf{q})$ element of the polarizability as a function of $q=|\mathbf{q}|$, for both \ch{hBN} and \ch{MoS2}. Its behavior in the long-wavelength limit $q \to 0$ is quadratic as expected. This feature is akin to 2D semiconductors and insulators (also 3D ones, not considered here), and it is universal~\cite{Thygesen_2017}, regardless of the details of the material as long as it presents a bandgap. From one material to another, what changes in the long-wavelength limit is the curvature at the origin (that depends on the direction of $\mathbf{q}$ if the material is anisotropic). We stress that the polarizability we compute is a strictly 2D one, in the sense that the sum over momentum in Eq.~\eqref{eq:Chi_0_static_results} is over in-plane momenta, as the BZ has no off-plane dimensions. The same can be said for the reciprocal lattice vectors, which have no $z$ component.

Since we have already discussed the results for the polarizability, we are now ready to analyze the dielectric function itself. This is the subject of the next section.

\subsection{Dielectric function}
\label{subsec:dielectric_function_results}

After the polarizability, we can compute the dielectric function to later study its inverse, which is needed to model the screened interacting potential. Each matrix element of the static dielectric function
is obtained through
\begin{equation}
\label{eq:dielectric_matrix_Results}
    \varepsilon_{\mathbf{G} \mathbf{G}'} (\mathbf{q}) = \delta_{\mathbf{G} \mathbf{G}'} - \sqrt{v_c(\mathbf{q} + \mathbf{G})} \chi^0_{\mathbf{G} \mathbf{G}'} (\mathbf{q}) \sqrt{v_c(\mathbf{q} + \mathbf{G}')}\,,
\end{equation}
where the quantity
\begin{equation}
\label{eq:2D_Coulomb_potential}    
    v_c (\mathbf{q}) = \frac{e}{2 \varepsilon_0 |\mathbf{q}| \Omega_\mathrm{\scriptscriptstyle UC}}
\end{equation}
is the 2D Fourier transform of the Coulomb potential, with $e$ being the elementary charge, $\varepsilon_0$ is the vacuum permittivity, and $\Omega_\mathrm{\scriptscriptstyle UC}$ the unit cell area. For reasons explained in the Methods~\ref{subsec:Chi&epsilon_Methods} section and also in section S.3 of the Supplementary Information (SI), we have adopted the symmetric representation for the dielectric function. Since what is needed for computing the screened Coulomb potential is the inverse dielectric matrix, we first need to calculate all elements of the direct matrix, and subsequently invert it.
In theory, the dielectric function in momentum space has infinite elements. In practice, we have to truncate it for a numerical treatment, since an analytical one is obviously impossible. 

Just as typically done in the literature in \textit{ab initio} calculations~\cite{BerkeleyGW_2012,GPAW_review_2024,Yambo_2009}, we truncate the dielectric matrix up to a certain maximum cutoff $G_{\mathrm{c}}^{\varepsilon}$ for the reciprocal lattice vectors. This means that we include all reciprocal lattice vectors $\mathbf{G}$ and $\mathbf{G}'$ such that $|\mathbf{G}| < G_{\mathrm{c}}^{\varepsilon}$ and $|\mathbf{G}'| < G_{\mathrm{c}}^{\varepsilon}$, for some value of $G_{\mathrm{c}}^{\varepsilon}$. Even though this cutoff has been explicitly mentioned in the literature in terms of energies, for example in Refs.~\onlinecite{BerkeleyGW_2012,Latini_Olsen_Thygesen_2015}, at the moment of writing this manuscript we are not aware of any reported study on the convergence of the inverse dielectric matrix elements with $G_{\mathrm{c}}^{\varepsilon}$ \markup{in the 2D case. Since it is not as well understood as its 3D counterpart~\cite{Hybertsen_Louie_1987},} it is instructive to analyze the convergence of the matrix elements $\varepsilon^{-1}_{\mathbf{G} \mathbf{G}'} (\mathbf{q})$ with the number of vectors $\mathbf{G}$ included. The detailed discussion of this issue is done in section~S.2 of the SI. In this work, and unlike in Ref.~\cite{BerkeleyGW_2012}, we do not have a single energy cutoff as a convergence parameter for truncating the band structure and the reciprocal lattice simultaneously. Instead, we include a sufficiently large number of empty bands $N_c$ for the polarizability matrix elements to converge, just as tested in Fig.~\ref{fig:Chi0_convergence}. Afterward, we use the $G_{\mathrm{c}}^{\varepsilon}$ parameter for the inclusion of the reciprocal lattice vectors, which is separate from $N_c$.
If it is converged with the number of conduction bands included in the sum of Eq.~\eqref{eq:Chi_0_static_results}, then, in principle, convergence of the dielectric matrix and its inverse with a high enough $G_{\mathrm{c}}^{\varepsilon}$ is attainable. However, the results shown in Fig.~S1 of the SI demonstrate that the convergence of the inverse dielectric matrix with increasing $G_{\mathrm{c}}^{\varepsilon}$ is quite slow, which calls for further investigation \markup{(see discussion at the end of section S.2 of the SI)}. Furthermore, the convergence of the exciton binding energy with $G_{\mathrm{c}}^{\varepsilon}$ must also be tested \markup{(see next subsection)}. However, as we will see next, choosing a small cutoff will be sufficient to compare our results with \textit{ab initio}~ones. 

\begin{figure*}[ht!]
    \centering
    \includegraphics[width=1.0\textwidth]{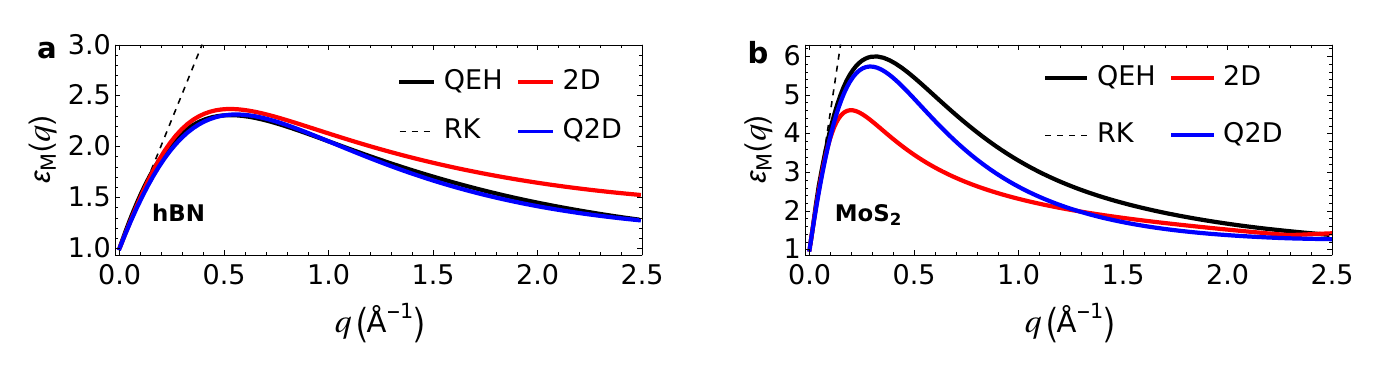}
    \caption[justification=justified]{\textbf{Macroscopic dielectric function.} Comparison between the effective 2D dielectric function obtained in this work with \textit{ab initio}~methods. Panels \pnl{a} and \pnl{b} display the results for \ch{hBN} and \ch{MoS2}, respectively. The \textit{ab initio} results were obtained using the QEH package for Python, which can be found in the literature~\cite{Andersen_Latini_Thygesen_2015}, and they are represented by the full black curves. The macroscopic dielectric functions obtained using the strictly 2D description developed in this work as per Eq.~\eqref{eq:macroscopic_epsilon_def_results} are represented in red. The blue curves represent the same quantity, but when obtained through the Q2D approach that we also introduce in this work, explained with detail in the Methods~\ref{subsec:Q2D_Methods} section. \markup{We applied the cutoff $G_{\mathrm{c}} = 3$\,{\AA}$^{-1}$ in both materials}. This accounts for seven reciprocal lattice vectors, including the null vector. We compare as well the full numerical $q$-dependence with the Rytova--Keldysh model in dashed black, with the $r_0$ parameter estimated from the small-$q$ linear regime in our numerical curves. For the Q2D results, we used for the monolayer thicknesses the values provided by the QEH package for each material.}
    \label{fig:macroscopic_2D_epsilon}
\end{figure*}

We discuss now the macroscopic effective dielectric function, which can be obtained from our 2D implementation of the microscopic one. For this, we employ the definition

\begin{equation}
\label{eq:macroscopic_epsilon_def_results}
    \varepsilon_{\mathrm{M}} (\mathbf{q}) = \frac{1}{\varepsilon^{-1}_{\mathbf{0} \mathbf{0}} (\mathbf{q})} \,,
\end{equation}

\noindent just as usually done in the literature~\cite{Huser_Olsen_Thygesen_2013,Latini_Olsen_Thygesen_2015,Thygesen_2017}.
The results are shown by the continuous red curves in Fig.~\ref{fig:macroscopic_2D_epsilon}. For comparison, we contrast our result with the macroscopic 2D dielectric function obtained from \textit{ab initio}~calculations in the literature represented by the continuous black curves. The data for those curves was obtained using the quantum-electrostatic heterostructure (QEH) model from Ref.~\onlinecite{Andersen_Latini_Thygesen_2015}, where the authors developed a Python package for estimating the dielectric properties of van der Waals structures. For the purpose of this manuscript, the QEH package was used for the particular case of monolayer \ch{hBN} and monolayer \ch{MoS2}, corresponding to panels~\ref{fig:macroscopic_2D_epsilon}\pnl{a} and~~\ref{fig:macroscopic_2D_epsilon}\pnl{b}, respectively.

It is important to note that the data obtained from the QEH Python package are based on previously done first-principles calculations, where the monolayer thickness is naturally incorporated. In the same figure, we present with a blue line the effective 2D dielectric function by taking the monolayer thickness into account. This is possible with the Q2D approach we developed in this work, which is an extension of the strictly 2D one for the dielectric function. In the Q2D approach, the dielectric function has a mixed $(\mathbf{q},z)$-representation, allowing for taking the off-plane degrees of freedom into account. In this way, the monolayer thickness can be included in our theory. For further technical details, we refer the reader to the Methods~\ref{subsec:Q2D_Methods} section.

It is worth noting the very good agreement between our curves (in red and blue) and the \textit{ab initio} ones (in black) in the low-$q$ limit. Our implementation constitutes then an efficient method of obtaining the small $q$-dependence of the effective 2D dielectric function, which is well captured using a linear model---the so-called Rytova--Keldysh model~\cite{Rytova_1967,keldysh2024coulomb}. The straight black lines in each panel correspond to the Rytova--Keldysh dielectric function for the respective material, which captures the close-to-zero $q$-dependence of the dielectric function as $\varepsilon_{\mathrm{RK}}(q) = 1 + r_0 q$ and its use is widely spread in the literature~\cite{Cudazzo_Tokatly_Rubio_2011,Quintela_Henriques_Tenorio_Peres_2022,Gomes_Trallero-Giner_Vasilevskiy_2021,Palacios.PhysRevB.91.245421}. The slope $r_0$ is obtained by fitting the straight curve to the numerical one, and is commonly denoted as screening length or screening parameter~\cite{Cudazzo_Tokatly_Rubio_2011}. It is also remarkable how well the Q2D approach captures the $q$-dependence of the macroscopic dielectric function in the entire $q$-range, especially for \ch{hBN}. Apart from small numerical discrepancies for \markup{$q\!\sim 0.3$\,{\AA}$^{-1}$ and $q\!\sim 1.8$\,{\AA}$^{-1}$}, the agreement between the black and blue curves is excellent. In the case of \ch{MoS2}, the strict 2D formalism underestimates the \textit{ab initio} screening in general, but approaches it for $q$ higher than $2.0$\,{\AA}$^{-1}$, sensibly. The Q2D also underestimates the \textit{ab initio} screening, but the improvement in the overall material dielectric response is striking.
The gain in using the Q2D approach over the 2D one is more evident for intermediate and higher values of momentum for both materials, namely after the linear approximation breaks down. Nonetheless, we can stress once again the excellent agreement among the three approaches for low values of $q$.

From the full dependence of the dielectric function, one can note that for $q \to \infty$ the macroscopic dielectric function approaches unity asymptotically, which is a characteristic of the macroscopic dielectric function in 2D (and also 3D) gapped materials~\cite{Trolle_Pedersen_Veniard_2017,Huser_Olsen_Thygesen_2013}.
Physically, this means that in the short-wavelength limit the spatial variations of the external potential occur so fast, that the system does not respond to the perturbation and test charges feel a total potential which is just the external potential averaged over a wavelength. With symbols, $\phi^{\text{total}} = \phi^{\text{ind}} + \phi^{\text{ext}} \approx \phi^{\text{ext}}$ for $q \to \infty$. This is in sharp contrast with the prediction of the Rytova--Keldysh model, for which the external potential is completely screened (due to an extra factor of $q$ in the denominator, making the total potential scale with $q$ as $\phi^{\text{total}} \sim 1/q^2$ instead of $\phi^{\text{total}} \sim 1/q$). The behavior of the \textit{ab initio} dielectric function when $q \to \infty$ is well captured by our methods, validating them once more.

One can notice in Fig.~\ref{fig:macroscopic_2D_epsilon} that the slope at the origin $r_0$ for \ch{hBN} differs significantly from the one for \ch{MoS2}.
This parameter can then be seen as an identity of each 2D semiconductor or insulator, as its value depends solely on the screening properties of the material.
As stated above, it is obtained from fitting a straight line to the 2D macroscopic dielectric function in the small $q$ limit, which one typically gets from first-principles calculations like DFT and DFT+$GW$ methods. However, such methods are typically 3D in nature, and the definition~\eqref{eq:macroscopic_epsilon_def_results} cannot be directly applied. First, one has to recover the $z$-dependence of the inverse dielectric function (for instance, with the inverse Fourier transform), and only then perform an off-plane average to obtain an effective 2D dielectric function. This process is usually a very heavy computational task.

Within our approach, we can easily estimate the screening parameter by a fit to the macroscopic dielectric function as per Eq.~\eqref{eq:macroscopic_epsilon_def_results}, \markup{both within the 2D and Q2D models, which give essentially the same $r_0$}. This has been confirmed by the results displayed in Fig.~\ref{fig:macroscopic_2D_epsilon}. We find $r_0 \approx 5.07$\,{\AA} for \ch{hBN}, and $r_0\approx 35.8$\,{\AA} for \ch{MoS2}. These values are in very good agreement with the literature.
We better show this agreement for \ch{hBN} in section~S.3 of the SI, where we plot our macroscopic dielectric function alongside two other dielectric functions obtained from \textit{ab initio} methods that can be found in the literature. The value obtained for \ch{MoS2}~is within the range of values $r_0 \approx 30-40$\,{\AA} provided by Ref.~\onlinecite{Berkelbach_Hybertsen_Reichman_2013}.
Different models may naturally lead to different estimates of this parameter. From inspection of the expression for the polarizability \eqref{eq:Chi_0_static_results}, in particular the denominator in the sum, one can expect that the bandgap will affect the outcome of our predictions.
Also, the estimated bandgap of a given material quantitatively differs from model to model.
This aspect is discussed in section~S.3 of the SI.
Interestingly, the dielectric function that we obtained using a hybrid functional for the DFT calculations is extremely close to the \textit{ab initio}~one from Ref.~\onlinecite{Latini_Olsen_Thygesen_2015}, where the band structure was obtained with LDA-DFT (which underestimates the true bandgap). Seemingly, the fine details of the employed Hamiltonian model are not relevant.

It is also important to point out that a more straightforward way of estimating the screening parameter is possible, by directly analyzing the low-$q$ behavior of the (head element of the) microscopic dielectric function. This has been pointed out by Li and Appelbaum in Ref.~\onlinecite{Li_Appelbaum_2019}, where they derived an expression for $r_0$ in terms of an in-plane momentum integral and momentum matrix element between conduction and valence bands, without involving any reciprocal lattice vectors. That expression allowed Quintela \emph{et~al.} to estimate in Ref.~\onlinecite{Mauricio_2024} the value of $r_0$ for the system therein studied, which was used in the same work for exciton calculations.

Once our approach has been validated by comparison with more sophisticated methods, the calculation of the excitonic energies follows as a natural application in the next section.

\subsection{Excitons}

We devote this section to a convergence analysis of the excitonic states, whilst using our numerical implementation of the screening in the calculation of the matrix elements of the Bethe--Salpeter Hamiltonian as per Eq.~\eqref{eq:Bethe_Salpeter_equation}.  In particular, we use the inverse dielectric matrix to compute the matrix elements of the interaction kernel $K$ as explained in the Methods~\ref{subsec:BSE_methods} section. \markup{We will limit our discussion to the results obtained using the strictly 2D dielectric matrix, making the necessary remarks regarding the Q2D approximation when appropriate}. In practice, we discard the exchange term and evaluate only the direct term $D$, so that $K=-D$. Regarding the singular terms that appear in the direct term when $\mathbf{q} = \mathbf{0}$, a proper treatment is necessary to avoid numerical errors. The Methods~\ref{subsec:BSE_methods} section discusses how the head term of the screened potential for zero momentum is regularized, since that is the only one that requires a non-trivial regularization scheme. In brief, we perform an average of the screened potential over a small circular region $\Omega_{\Gamma}$ around the origin, just like expressed in Eq.~\eqref{eq:head_term_regularization}. In section~S.3 of the SI we argue why the singular terms of the form $W_{\mathbf{G} \mathbf{0}}$ and $W_{\mathbf{0} \mathbf{G}}$, called wing terms, can simply be set to zero.
The terms of the form $W_{\mathbf{G} \mathbf{G}'}$ where $\mathbf{G} \neq \mathbf{0}$ and $\mathbf{G}' \neq \mathbf{0}$, called body terms in the literature, do not need any regularization procedure.

At this stage, a study on how the excitonic binding energies behave with increasing refinement of the BZ, $N$, and increasing cutoff for the dielectric function, $G_{\mathrm{c}}^{\varepsilon}$, is a natural exercise to do. Due to practical implementations of the singular head element and the double sum over reciprocal lattice vectors for $D$ in Eq.~\eqref{eq:direct_interaction_term}, we have two extra parameters to account for: (i) the radius of the regularization region $\Omega_{\Gamma}$, that we denote by $q_0$; (ii) a cutoff for the reciprocal lattice vectors when calculating $D$, that we can denote by $G^X_\mathrm{c}$.
As the parameter space is big, we make two separate convergence studies where we focus on the excitonic ground state. First, we study the convergence of the exciton energy level with $q_0$ and $N$ fixing a certain value for the dielectric matrix cutoff $G_{\mathrm{c}}^{\varepsilon}$. This time, we include all the reciprocal vectors in $D$, i.e., we set $G^X_\mathrm{c} = G_{\mathrm{c}}^{\varepsilon}$. Second, we fix certain $q_0$ and $N$ to study the convergence of the exciton binding energy with $G_{\mathrm{c}}^{\varepsilon}$ and $G^X_\mathrm{c}$.

\begin{figure*}[ht!]
    \centering
    \includegraphics[width=1.00\linewidth]{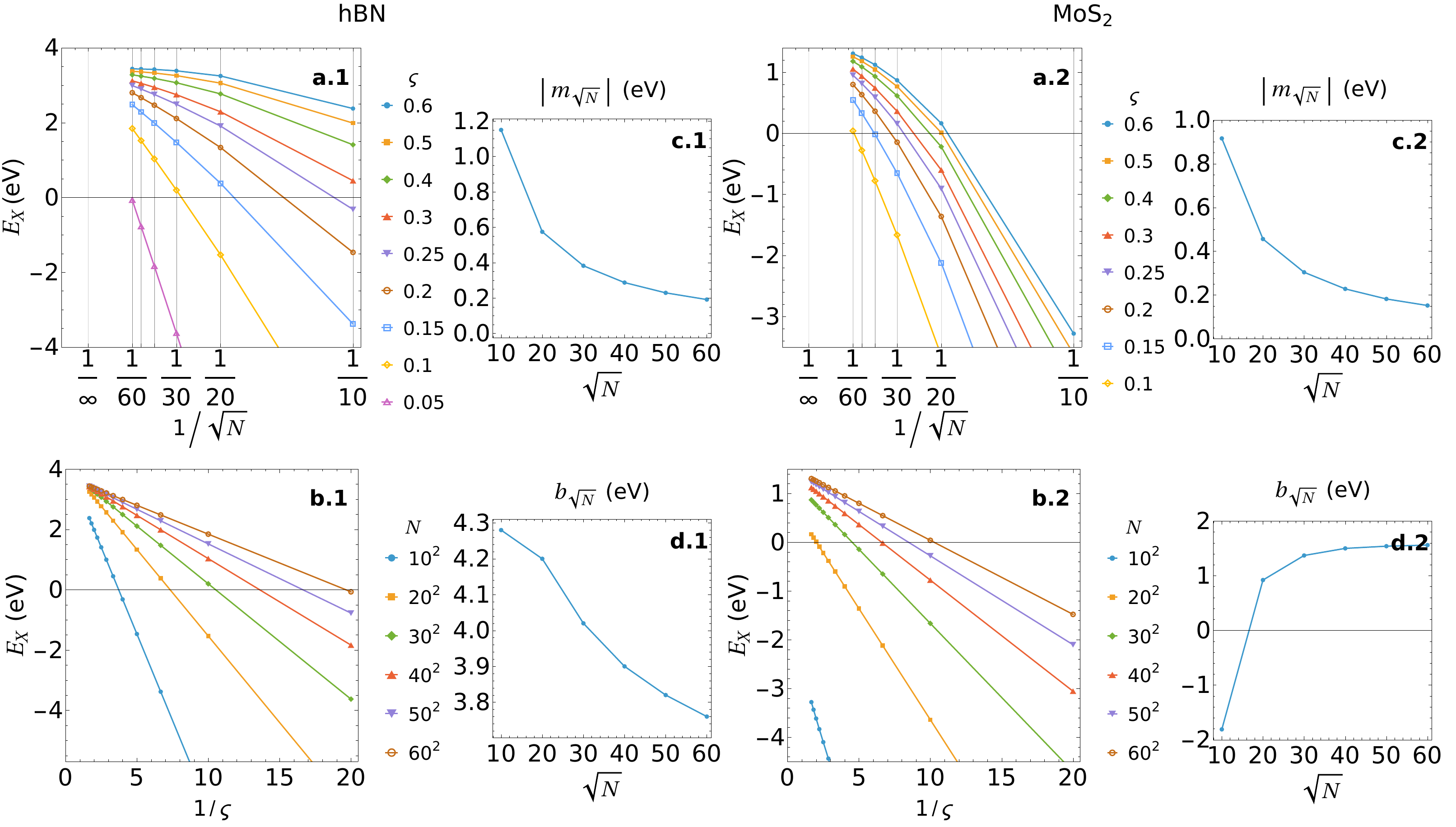}
    \caption{\textbf{Convergence of the \markup{excitonic energy} with the BZ mesh size and regularization radius.} Panels \pnl{a.1}-\pnl{d.1} present the results pertaining to hBN, while panels \pnl{a.2}-\pnl{d.2} pertain to \ch{MoS2}. Panels \pnl{a.1} and \pnl{a.2} show the ground state excitonic energy level $E_X$ as a function of $N$ for different realizations of $\varsigma$, which defines the radius of the regularization region as $q_0=\varsigma k_0$, $k_0$ being the length of the momentum vector closest to the origin in the BZ. Panels \pnl{b.1} and \pnl{b.2} show the dependence of $E_X$ on $\varsigma^{-1}$ for different realizations of $N$, which follows a linear model as $E_X = m_N \varsigma^{-1} + b_N$. The plot in panels \pnl{c.1} and \pnl{c.2} shows the behavior of the slope $m_N$ with $N$, whereas panels \pnl{d.1} and \pnl{d.2} plot the origin off-set $b_N$ versus $N$. \markup{For the hBN calculations, we used $G_{\mathrm{c}}^{\varepsilon} = 3$~\AA$^{-1}$, while for \ch{MoS2} we used $G_{\mathrm{c}}^{\varepsilon} = 5$~\AA$^{-1}$}.}
    \label{fig:Exciton_convergence_hBN_&_MoS2}
\end{figure*}

The results of the first convergence study are displayed in Fig.~\ref{fig:Exciton_convergence_hBN_&_MoS2} in two sets of four panels, the left ones \pnl{a.1} to \pnl{d.1} pertaining to hBN and the right ones \pnl{a.2} to \pnl{d.2} pertaining to \ch{MoS2}. For hBN we set $G_{\mathrm{c}}^{\varepsilon} = 3$\,{\AA}$^{-1}$, while for \ch{MoS2} we set $G_{\mathrm{c}}^{\varepsilon} = 5.1$\,{\AA}$^{-1}$. In panels \pnl{a.1} and \pnl{a.2} we plot the energy level of the lowest exciton, denoted by $E_X$, as a function of the number of $\mathbf{k}$ points in the BZ, $N$, for different realizations of $q_0$. In each realization, $q_0$ is given by a fraction $\varsigma$ of the length of the momentum vector closest to the origin, which we denote by $k_0$. In both materials, the convergence is slow with $N$, with \ch{MoS2} being the worst case.
However, plotting $E_X$ as a function of $1/\varsigma$ for different realizations of $N$, as shown in panels \pnl{b.1} and \pnl{b.2}, shows that we can model the dependence $E_X(\varsigma,N)$ with a linear model whose parameters vary among different values of $N$. We can thus write $E_X(\varsigma,N) = m_N \varsigma^{-1} + b_N$, where $m_N$ is the slope and $b_N$ the origin off-set for a given $N$. It is evident in panels \pnl{b.1} and \pnl{b.2} that as we increase $N$, the straight line shifts upwards, and the slope decreases in absolute value. In the limit $N \to \infty$ one expects that the line becomes horizontal, meaning that the exciton does not depend on the regularization. This makes sense, since the regularization of the singular terms in the screened potential is a numerical issue---in the limit of an infinite number of $\mathbf{k}$ points the sum over momentum in Eq.~\eqref{eq:Bethe_Salpeter_equation} is an integral, and removing a single $\mathbf{k}$ point in the integration region bears no difference in the final result. Then, the true excitonic energy is the limit $\lim_{N \to \infty} b_N$. To test our hypothesis, we fit a linear model to each curve in panels \pnl{b.1} and \pnl{b.2}, and obtain the values of the parameters $m_N$ and $b_N$ for every realization of $N$. Those values are displayed in panels~\pnl{c} and \pnl{d} of each material, respectively.
In panels~\pnl{c.1} and \pnl{c.2}, we can clearly see that $\lim_{N \to \infty} m_N = 0$ holds for both hBN and \ch{MoS2}. Discussing now the origin off-set, in panel~\pnl{d.1} we see that $b_N$ decreases as $N$ grows. In panel~\pnl{d.2} we can see that having $N > 60$ would not bring a significant difference in the value of $b_N$ for \ch{MoS2}. As a side note, in all the fits that we have performed, the value of $r^2$ is approximately 1 up to a rounding digit in the fifth decimal place, if not exactly 1.

Continuing our analysis, we take as an estimate for the excitonic energies the values of $b_{\sqrt{N}=60}$ for both materials. For hBN, this means $E_X \approx b_{\sqrt{N}=60} = 3.76$\,eV \mymarkup{(see Fig.~5\pnl{d}.1)}, while for \ch{MoS2} we have $E_X \approx b_{\sqrt{N}=60} = 1.56$\,eV \mymarkup{(see Fig.~5\pnl{d}.2)}. Considering the values of the bandgaps for both materials given in the legend of Fig.~\ref{fig:hBN_&_MoS2_bands}, the resulting binding energy for hBN is $E_b = 2.32$\,eV. This value is not too distant from \textit{ab initio}~ones found in the literature, for instance $2.05$\,eV in Ref.~\onlinecite{Latini_Olsen_Thygesen_2015}, $1.81$\,eV in Ref.~\onlinecite{PhysRevB.108.075413}, $2.08$\,eV in Ref.~\onlinecite{Zhang_Ong_Ruan_Wu_Shi_Tang_Louie_2022}, or, more recently, $1.88$\,eV in Ref.~\onlinecite{Toni_Juanjo_2025}. In the case of \ch{MoS2}, we obtain the binding energy $E_b  = 0.53$\,eV \markup{using the exciton energy estimated in the beginning of this paragraph}, which is also comparable with \textit{ab initio}~results. For example, the authors of Ref.~\onlinecite{Latini_Olsen_Thygesen_2015} obtained therein the value $E_b=0.43$\,eV. However, an experimental value of $0.44$\,eV was reported in Ref.~\onlinecite{Hill_Rigosi_Roquelet_2015} \markup{for a \ch{MoS2} monolayer deposited on silica. Therefore, the real value should be higher than $0.44$\,eV, and has been estimated to be $0.63$~eV within first-principles~\cite{Diana_Jornada_Louie}, not too distant from our estimated value of $0.53$~eV}. As a disclaimer, the analysis performed hitherto was done for a singular value of the cutoff $G_\mathrm{c}^{\varepsilon}$ and the strict 2D dielectric screening \markup{(as shown below, the results from the Q2D screening can only improve the agreement in this regard)}.

We now compute the excitonic energies while varying the cutoff $G_{\mathrm{c}}^{\varepsilon}$, which determines the size of the inverse dielectric matrix incorporated in the interaction matrix element $D$ in Eq.~\eqref{eq:direct_interaction_term}, and the cutoff $G^X_{\mathrm{c}}$, which determines the number of terms that appear in the double sum over reciprocal lattice vectors that composes $D$. The focus is again the exciton with lowest energy. The results are displayed in Table~\ref{tab:Exciton_binding_energy} below. In it, we can identify some patterns worth noting. First, if we consider the first row of each table, for which $G_{\mathrm{c}}^X = 0$, and reading the values from left to right, we can see that as $G_{\mathrm{c}}^{\varepsilon}$ increases, the binding energy also increases. This can be explained by understanding what happens to the 2D effective dielectric function as $G_{\mathrm{c}}^{\varepsilon}$ increases. It can be seen in Fig.~S4 of the SI that the macroscopic dielectric function is overestimated if one includes too few reciprocal lattice vectors in its calculation. Then, as we increase $G_{\mathrm{c}}^{\varepsilon}$ and include more vectors $\mathbf{G}$, as $\varepsilon_{\mathrm{M}}(\mathbf{q})$ overall decreases, the screened potential given by $v_c(\mathbf{q})/\varepsilon_{\mathrm{M}}(\mathbf{q})$ gets stronger. The attraction between the electron and the hole intensifies, translating into a higher binding energy. Eventually, it grows bigger than the electronic bandgap, which is unphysical (occurs for the entries in the tables with $> \Delta$). To draw conclusions with physical meaning, we must include all $\mathbf{G}$ vectors when evaluating $D$. Doing so, we obtain the results in the diagonal cells of each table. Even though the matrix elements of the inverse dielectric function display a very slow convergence as discussed in section S2, the binding energy of the exciton seems to converge faster than the inverse dielectric function. Despite this apparent convergence, the possibility of achieving a satisfying convergence in practice is not evident.  This slow convergence may be due to the slow decay of the Coulomb potential \mymarkup{in the strict 2D approach}, which conditions the convergence of the dielectric matrix \mymarkup{and of $D$}. Regarding the values obtained using the Q2D approach, that are shown in Table~\ref{tab:Exciton_binding_energy} along the diagonal cells inside brackets, the convergence is substantially improved and, most likely, \mymarkup{something similar would be observed} in bulk 3D materials, where \mymarkup{the Fourier transform of the Coulomb potential goes as $v_c(\mathbf{q}) \sim 1/q^2$. This dependency can be recovered in the Q2D case by taking the short-wavelength limit in Eq.~(S.26) of the SI}. 

Regardless of the inverse dielectric matrix being converged or not, one has to include enough $\mathbf{G}$ vectors \mymarkup{in number and up to a certain cutoff} to keep the symmetry of the interaction and to obtain the expected degeneracies. \mymarkup{Naturally, the smallest possible cutoff is system dependent.} This issue is more delicate in hBN than in \ch{MoS2} because \mymarkup{in} the former \mymarkup{the screening is less efficient, whereas the interaction strength decays faster in \ch{MoS2}. Furthermore, the fact that the chemical bond in hBN is essentially ionic leads to a more inhomogeneous electron density than in \ch{MoS2}, where the bond is covalent. Hence, we need to include more reciprocal lattice vectors in the calculations for hBN. We verified that for \ch{MoS2} setting $G^X_{\mathrm{c}} = 0$ does not have a big impact in the expected degeneracies (a difference smaller than 0.003 eV for the lowest excitons), while for hBN one has to include all momenta up to the third BZ to obtain degenerate exciton states}. 

\begin{table*}[ht!]
    \caption{\textbf{Convergence of the exciton binding energy with the cutoff.} These tables examine the convergence of the ground state excitonic binding energy, for \ch{hBN} on the left and for \ch{MoS2} on the right, with the cutoff $G_{\mathrm{c}}^{\varepsilon}$ for the dielectric matrix, and with the cutoff $G_{\mathrm{c}}^X$ for the interaction matrix elements (always with $G_{\mathrm{c}}^X < G_{\mathrm{c}}^{\varepsilon}$) \mymarkup{when using the 2D scheme for the screening. The values inside curved brackets along the diagonals were obtained within the Q2D scheme}. We have used $N=60^2$, $N_c=N_v=1$, and we have excluded the exchange interaction term in both cases. For the size of the regularization region, we used the radius $q_0 = \varsigma k_0$, where $k_0$ is the norm of the wavevector(s) closest to the origin. We chose $\varsigma=0.6$ for both materials. To obtain the binding energies in eV, we used for the electronic bandgap of \ch{hBN} $\Delta =6.08$\,eV, and for the bandgap of \ch{MoS2} we used $\Delta = 2.08$\,eV, both values extracted from the band structure of these materials in Fig.~\ref{fig:hBN_&_MoS2_bands}. The entries in each table with $> \Delta$ are entries for which we have obtained a binding energy higher than the electronic bandgap. As this is not physical, we just omit the values obtained.}
    \centering
    \begin{minipage}{0.5\linewidth}
        \centering
        \begin{tabular}{cc||cccccc}
            \cline{3-8}
            &   & \multicolumn{6}{c|}{$G^{\varepsilon}_{\mathrm{c}}$  (\AA$^{-1}$)}                                            \\ \cline{3-8}
            &                                                & 0                          & 3                & 5.1              & 6              & 8                      & \multicolumn{1}{c|}{9}                       \\ \cline{8-8} \hline \hline
            \multicolumn{1}{|c|}{\multirow{10}{*}{$G^X_{\mathrm{c}}$}} & 0 & \multicolumn{1}{c|}{\phantom{0.000}} & 4.209               & 5.617                      & 5.814                  & 6.071               & $ > \Delta $        \\
            \multicolumn{1}{|c|}{}                           & 3 & \multicolumn{1}{c|}{\phantom{X}} & \begin{tabular}{@{}c@{}} 2.563 \\ \markup{(2.185)}\end{tabular}               & 4.440                    & 3.863                 & 3.872               & 4.386          \\ \cline{4-4} \cline{4-4}
            \multicolumn{1}{|c|}{}                           & 5.1 & \phantom{X}                    & \multicolumn{1}{c|}{\phantom{X}} & \begin{tabular}{@{}c@{}}  2.857 \\ \markup{(2.247)} \end{tabular}                       & 3.054                  & 3.948               & 4.690          \\ \cline{5-5} \cline{5-5}
            \multicolumn{1}{|c|}{}                           & 6 & \phantom{X}                      & \phantom{X}                      & \multicolumn{1}{c|}{\phantom{X}}      & \begin{tabular}{@{}c@{}}  2.894 \\ \markup{(2.237)} \end{tabular}                  & 3.542               & 4.391          \\ \cline{6-6} \cline{6-6}
            \multicolumn{1}{|c|}{}                           & 8 & \phantom{X}                      & \phantom{X}                      & \phantom{X}                           & \multicolumn{1}{c|}{\phantom{X}}   & \begin{tabular}{@{}c@{}}  2.961 \\ \markup{(2.238)} \end{tabular}               & 5.872          \\ \cline{7-7}
            \multicolumn{1}{|c|}{}                           & 9 & \phantom{X}                      & \phantom{X}                      & \phantom{X}                           & \phantom{X}                        & \multicolumn{1}{c|}{\phantom{X}} & \begin{tabular}{@{}c@{}}  3.078  \\ \markup{(2.256)} \end{tabular}\\  \cline{1-2} \cline{1-2} \cline{8-8}
        \end{tabular}
    \end{minipage}%
    \begin{minipage}{0.5\linewidth}
        \centering
        \begin{tabular}{cc||cccccc}
                \cline{3-8}
                &   & \multicolumn{6}{c|}{$G^{\varepsilon}_{\mathrm{c}}$ (\AA$^{-1}$)}                                            \\ \cline{3-8}
                &   & 0                      & 3                      & 4                           & 5                        & 7                      & \multicolumn{1}{c|}{8}                       \\ \cline{8-8} \hline \hline
                \multicolumn{1}{|c|}{\multirow{10}{*}{$G^X_{\mathrm{c}}$}} & 0 & \multicolumn{1}{c|}{\phantom{0.000}} & 0.980               & 1.568                    & 1.401                 & 1.952               & $ > \Delta $        \\
                \multicolumn{1}{|c|}{}                           & 3 & \multicolumn{1}{c|}{\phantom{X}} & \begin{tabular}{@{}c@{}}0.756 \\ \markup{(0.7066)}\end{tabular}               & 1.619                    & 1.786                 & $ > \Delta $                    & $ > \Delta $          \\ \cline{4-4} \cline{4-4}
                \multicolumn{1}{|c|}{}                           & 4 & \phantom{X}                      & \multicolumn{1}{c|}{\phantom{X}} & \begin{tabular}{@{}c@{}}0.775 \\ \markup{(0.7088)} \end{tabular}                   & 1.106                 & $ > \Delta $                    & $ > \Delta $          \\ \cline{5-5} \cline{5-5}
                \multicolumn{1}{|c|}{}                           & 5 & \phantom{X}                      & \phantom{X}                      & \multicolumn{1}{c|}{\phantom{X}}      & \begin{tabular}{@{}c@{}}0.778 \\ \markup{(0.7093)}\end{tabular}               & $ > \Delta $                    & $ > \Delta $          \\ \cline{6-6} \cline{6-6}
                \multicolumn{1}{|c|}{}                           & 7 & \phantom{X}                      & \phantom{X}                      & \phantom{X}                           & \multicolumn{1}{c|}{\phantom{X}}   & \begin{tabular}{@{}c@{}}0.792 \\ \markup{(0.712)} \end{tabular}               & 1.301          \\ \cline{7-7}
                \multicolumn{1}{|c|}{}                           & 8 & \phantom{X}                      & \phantom{X}                      & \phantom{X}                           & \phantom{X}                        & \multicolumn{1}{c|}{\phantom{X}} & \begin{tabular}{@{}c@{}}0.799 \\ \markup{(0.713)} \end{tabular} \\  \cline{1-2} \cline{1-2} \cline{8-8}
            \end{tabular}
    \end{minipage}
    \label{tab:Exciton_binding_energy}
\end{table*}

\mymarkup{Our findings are} in agreement with results previously reported in the literature~\cite{Ridolfi_2020}. We have included Tables S3 and S4 in the SI, displaying the first excitonic energies, binding energies and respective degeneracies for hBN and \ch{MoS2}, respectively, using the Q2D screening. For this, we have chosen a single cutoff from Table~\ref{tab:Exciton_binding_energy} for each material and used the same input parameters. As a small disclaimer, we have obtained the degeneracies for \ch{MoS2} that we expected with our model to validate our methodology, but the model is not realistic. If one wants to better compare with experiments, including SOC in the material Hamiltonian is strictly necessary, which would lift the degeneracies we obtain.

\markup{We verified that when performing exciton calculations in reciprocal space using the first version of the code with the Rytova--Keldysh potential~\cite{Alex_Xatu_2024}, the binding energies come out slightly underestimated with the respect to the results obtained using our implemented dielectric function. This is a consequence of the Rytova--Keldysh model overestimating the screening, thus underestimating the electron--hole attraction strength. Whence, one would expect an increase in the oscillator strength, which we confirmed by computing the (real part of the) optical conductivity using the XATU code. Naturally, we obtained enhanced resonant peaks compared to those obtained from using the Rytova--Keldysh screening. To see this effect, the reader can consult Fig.~S7 at the~end~of~the~SI.}

\section*{Discussion}
\label{sec:conclusions}

\markup{We have developed an efficient atomistic framework that bridges analytical models~\cite{Trolle_Pedersen_Veniard_2017,Cudazzo_Tokatly_Rubio_2011} and \textit{ab initio} descriptions of screening~\cite{Hybertsen_Louie_1987} and excitons for 2D materials~\cite{Louie_2000}. Exploiting the reduced dimensionality, our strictly 2D formalism provides a microscopically detailed dielectric matrix (and its inverse) at a fraction of the computational cost of first-principles methods.  We have benchmarked the method on hBN and MoS$_2$ monolayers, accurately capturing both long- and short-wavelength screening. Extending the approach to a quasi-2D (Q2D) description (see the Methods~\ref{subsec:Q2D_Methods} section) clarifies the validity range of the zero-thickness approximation and improves agreement with \textit{ab initio} dielectric responses, all while remaining computationally inexpensive.}

\markup{The agreement with the \textit{ab initio} results is remarkable, given the simplified nature of our treatment, in the sense that two approximations were made.
First, the point-like orbital approximation for the plane-wave matrix elements (explained in detail in Methods~\ref{subsec:Chi&epsilon_Methods}), which does not present any setback.
This approximation is advantageous to evaluate the polarizability as per Eq.~\eqref{eq:Chi_0_static_results}, in which the numerator inside the sum can be computed with negligible computational effort.
On the contrary, in \textit{ab initio}~approaches where a reciprocal space description of screening is adopted, these matrix elements are typically evaluated through FFTs, implying computational and implementation complications that we avoid altogether.
Second, in the strict 2D approach we have ignored completely the physics pertaining to the off-plane degrees of freedom. As seen in Fig.~\ref{fig:macroscopic_2D_epsilon}, this approximation, which is discussed more extensively in the Methods section, does not influence the low-$q$ dielectric properties of the material. Therefore, one can readily recover the Rytova--Keldysh dielectric function while estimating the screening parameter with an accuracy akin to first-principles approaches.
For intermediate and high values of $q$, high enough so that we no longer stand in the linear-in-$q$ regime, the strict 2D approach overestimates the screening for hBN and underestimates it for \MoS~when compared to \textit{ab initio} results with similar cutoffs. Fortunately, this penalty in the screening can be fixed by incorporating the non-vanishing monolayer thickness through the Q2D approach for the dielectric function. In this way, our approach generalized to the dynamic and metallic cases can possibly bring new insights into the fields of plasmonics and photonics in low-dimensional materials.}

\markup{Although our screening methodology has been validated  against  \textit{ab initio} methods, there are no experimentally measured macroscopic static dielectric functions to compare with. However,
using the resulting dielectric matrix as input for BSE calculations implemented in the XATU code, we also obtained exciton energies consistent with first-principles results, which can be compared with experiments as well. \mymarkup{On the one hand, the exciton energies obtained with the strict 2D approach overestimate the \textit{ab initio} ones. On the other hand, the Q2D calculations bring the estimated binding energies down. Even though the 2D description is quite direct, the zero-thickness approximation is limiting and the Q2D description should always be used instead.} The efficiency of the \mymarkup{approaches} enables detailed convergence studies, including refinement of the Brillouin zone, integration cutoffs, and local-field effects in the BSE kernel, which are often computationally demanding to be carried out in fully \textit{ab initio} treatments.}

\section*{Methods}
\label{sec:Methods}
\setcounter{subsection}{0}
\subsection{DFT calculations}
\label{subsec:DFT&GW_Methods}

To obtain the band structure and the model Hamiltonian of the materials in consideration we performed DFT calculations with the Kohn--Sham method~\cite{Cohen_Louie_2016,Martin_2004,Kohanoff_2006}.
In the early days of DFT, the electron--electron exchange and correlation contribution to the energy of the system was modeled within the LDA for the exchange--correlation functional. The LDA picture is well-known for underestimating the true electronic bandgap of semiconductors and insulators; this is the fundamental gap problem.
To correctly estimate the bandgap of semiconductors and insulators, one can perform $GW$ calculations as a post-processing step~\cite{Martin_Reining_Ceperley_2016}. However, these involve the evaluation of the dielectric function beyond the static limit, which can be very heavy numerically. 

As an alternative, one can introduce hybrid functionals to better model exchange--correlation effects.
\markup{Hybrid functionals combine the exact and unscreened Fock exchange with an approximated screened one}, say at the LDA or at the generalized gradient approximation (GGA) level~\cite{Martin_2004,Kohanoff_2006}. Generically, the hybrid functionals permit to better estimate the electronic bandgap than non-hybrid ones when compared with the experimental values~\cite{PhysRevMaterials.8.043803}, and at a much lower computational cost than $GW$ methods.
In this work, we bypass computationally intensive $GW$ calculations by using the functional HSE06~\cite{Heyd_Scuseria_Ernzerhof_2003}, which is a \markup{hybrid, range-separated functional}~\cite{Perdew_1996}.
The calculations were performed on a basis of localized Gaussian orbitals with the CRYSTAL code~\cite{CRYSTAL_2023}, as explained in the Results~\ref{subsec:Band_structure_results} section.  Alternatively,  a Wannier interpolation can be performed after obtaining the quasiparticle band structure $\{\epsilon_{n\mathbf{k}}\}$ through DFT calculations using a basis of delocalized functions (plane waves, for instance). Both allow for obtaining a Hamiltonian in a basis of localized orbitals, which is essential for us to apply the point-like orbital approximation. First mentioned in the Results~\ref{subsec:Results_Polarizability} section, this approximation shall be technically explained in the next section.
The Hamiltonian obtained for each material serves then as a basis for describing the excitonic problem within an interacting picture beyond the mean-field level. The screening calculations necessary for the excitonic problem are performed on top of the same basis and are explained in the following subsections. Finally, the dielectric screening problem is followed by the calculation of the Bethe--Salpeter Hamiltonian and its diagonalization.

\subsection{Dielectric properties: polarizability and dielectric function}
\label{subsec:Chi&epsilon_Methods}

The non-interacting irreducible polarizability in momentum--frequency space reads~\cite{Adler_1962,Cohen_Louie_2016}

\begin{widetext}
\begin{equation}\label{eq:Chi_0_def}  
    \chi^0_{\mathbf{G} \mathbf{G}'} (\mathbf{q};\omega) = \frac{1}{N} \sum_{n n',\mathbf{k}\sigma} \frac{f_{n \mathbf{k}} - f_{n' \mathbf{k} + \mathbf{q}}}{\epsilon_{n \mathbf{k}} - \epsilon_{n' \mathbf{k} + \mathbf{q}} + \hbar \omega + \mathrm{i} \hbar \alpha}  \big\langle n,\mathbf{k} \big| \mathrm{e}^{-\mathrm{i}(\mathbf{q} + \mathbf{G}) \cdot \mathbf{r}} \big| n',\mathbf{k} + \mathbf{q} \big\rangle  \big\langle n',\mathbf{k} + \mathbf{q} \big| \mathrm{e}^{\mathrm{i}(\mathbf{q} + \mathbf{G}') \cdot \mathbf{r}} \big| n,\mathbf{k} \big\rangle\,,
    \end{equation}
\end{widetext}
%
where $| n, \mathbf{k} \rangle$ represents a single-particle Bloch state of an electron (or hole) in the band $n$ with momentum $\mathbf{k}$, and $f_{n \mathbf{k}}$ is the occupation factor of the same state, $N$ is the number of $\mathbf{k}$ points or unit cells, so that the volume (area actually) of the crystal is given by $\Omega = N \Omega_\mathrm{\scriptscriptstyle UC}$, with $\Omega_\mathrm{\scriptscriptstyle UC}$ being the volume of the unit cell, and $\alpha \to 0$ is an infinitesimal positive parameter.

The plane-wave matrix elements of the type $\langle n,\mathbf{k} | \mathrm{e}^{-\mathrm{i}(\mathbf{k}' - \mathbf{k} + \mathbf{G}) \cdot \mathbf{r}} | n',\mathbf{k}' \rangle \equiv I^{\mathbf{G}}_{n\mathbf{k}, n' \mathbf{k}'}$ are evaluated along the 2D material, ignoring the off-plane dimension so that the position vector $\mathbf{r}$ is strictly in-plane.
Within the linear combination of atomic orbitals (LCAO) and tight-binding (TB) approximations, the wavefunctions $\psi_{n \mathbf{k}} (\mathbf{r}) = \langle \mathbf{r} | n, \mathbf{k} \rangle$ of the Bloch states, in the lattice gauge, are written in terms of localized orbitals $\phi_{\alpha}$ as 
\begin{equation}
\label{eq:Bloch_state_LCAO}
    \psi_{n \mathbf{k}} (\mathbf{r}) = \frac{1}{\sqrt{N}} \sum_{\bR} \mathrm{e}^{\mathrm{i} \mathbf{k} \cdot \bR} \sum_{i \alpha} C_{i \alpha}^{n \mathbf{k}} \phi_{\alpha} (\mathbf{r} - \bR - \mathbf{t}_i)\,,
\end{equation}
with the sum in $\bR$ being over (all $N$) unit cells of the Bravais lattice, and $\mathbf{t}_i$ designates the position of the $i$th atom within the unit cell.
The coefficients $C_{i \alpha}^{n \mathbf{k}} = (\bC^{n \mathbf{k}})_{i \alpha}$ are commonly called the TB coefficients, and they form the eigenvectors $\bC^{n \mathbf{k}}$ of the Bloch Hamiltonian, in the lattice gauge defined by $H(\mathbf{k}) = \sum_{\bR} \exp(\mathrm{i} \mathbf{k} \cdot \bR ) H(\bR)$, where the Fock matrices $H(\bR)$ are obtained from the DFT calculations.  Mathematically, in a basis where the orbitals are orthogonal, we write $H(\mathbf{k}) \bC^{n \mathbf{k}} = \epsilon_{n\mathbf{k}} \bC^{n \mathbf{k}}$.

In equilibrium, the occupation factor $f_{n \mathbf{k}}$ follows simply the Fermi--Dirac distribution, which in the zero absolute temperature limit is just a step function. It means that for an energy below the Fermi energy $f_{n \mathbf{k}} = 1$, and zero otherwise. In this paper we discuss only intrinsic semiconductors and insulators, that at zero temperature have all the bands below the Fermi energy completely filled, and those above the Fermi energy are empty. The ground state is that of all the valence bands filled and the conduction bands empty, with a gap of forbidden energies separating them. In the static limit $\omega\to0$ and at zero temperature, and if time-reversal symmetry (TRS) is present, the expression~\eqref{eq:Chi_0_def} above reduces to expression~\eqref{eq:Chi_0_static_results}. We refer to section~S.1 of the SI for a detailed proof. The matrix elements in the numerator can be computed by evaluating the real-space integral
\begin{equation}
\label{eq:plane_wave_mel}
\begin{split}
    I^{\mathbf{G}}_{n\mathbf{k}, n'\mathbf{k}'} &= \int_\Omega \psi^*_{n\mathbf{k}} (\mathbf{r})\, \mathrm{e}^{-\mathrm{i}(\mathbf{k}' - \mathbf{k} + \mathbf{G}) \vdot \mathbf{r}} \psi_{n' \mathbf{k}'} (\mathbf{r}) \,\mathrm{d}\mathbf{r}  \\
    &= \sum_{i \alpha} (C_{i \alpha}^{n \mathbf{k}})^* C_{i \alpha}^{n' \mathbf{k}'}  \mathrm{e}^{-\mathrm{i}(\mathbf{k}' - \mathbf{k} + \mathbf{G})  \vdot \mathbf{t}_i} 
\end{split}
\end{equation}
taken over the entire crystal, and where the last equality was obtained after replacing the Bloch state wavefunctions $\psi_{n\mathbf{k}} (\mathbf{r})$ by their respective LCAO expansion as per \eqref{eq:Bloch_state_LCAO} and simplifying. We also used the identity $\phi_{\alpha} (\mathbf{r} - \bR - \mathbf{t}_{i}) \phi_{\beta} (\mathbf{r} - \bR' - \mathbf{t}_j) \approx \delta_{\alpha \beta} \delta_{\bR \bR'} \delta_{ij} \delta(\mathbf{r} - \bR - \mathbf{t}_i)$, the so called point-like orbital approximation within which the orbitals are extremely localized around the atomic positions~\cite{Alex_Xatu_2024}.
We make here a pertinent technical remark. These orbitals at hand might not necessarily be atomic orbitals in a TB description of the Hamiltonian, but they may also be interpreted as Wannier orbitals~\cite{Blount_1962} of a Hamiltonian obtained from the Wannier interpolation of the DFT bands.
So, instead of seeing these orbitals as several atomic orbitals localized at each position $\mathbf{t}_i$ of the $i$th atom in the motif of the unit cell, we can see them as a set of Wannier orbitals, each one localized at a Wannier center that in general does not coincide with atomic sites in the lattice. Then, $\mathbf{t}_i$ is no longer a position of an atom in the motif, but the $i$th Wannier center within the unit cell.

We have tested our implementation with Wannier models as well and with good results, in particular for \ch{MoS2}, something discussed in more detail in section S.4 of the SI. During our investigations, we observed that the head element of the polarizability for an isotropic material when plotted as a function of $q=|\mathbf{q}|$ with $\mathbf{q}$ covering the whole BZ, like in Fig.~\ref{fig:Chi0_vs_q_hBN_&_MoS2}, may present considerable numerical fluctuations. Instead, one should obtain a dependence on $q$ presenting a single curve as expected for isotropic materials.
Those models, with the band structure calculated through DFT followed by a Wannier interpolation using the \textsc{Wannier90} code, were obtained without forcing the symmetry preservation when running \textsc{Wannier90}.
Therefore, the symmetries of a material model Hamiltonian may play a role in its dielectric properties, most likely those related to the space group of the crystalline material, which can be violated when performing a faulty Wannier interpolation.
However, the exact role that space group symmetries play in the polarizability is unclear to us.

The irreducible polarizability $\chiirr$ and the full interacting one $\chi$ are connected through the Dyson equation $\chi = \chiirr + \chiirr V \chi$~\cite{Bruus_Flensberg_2004,Martin_Reining_Ceperley_2016}, where $V$ is the Coulomb interaction kernel, and in general includes the direct unscreened Coulomb interaction $v_c(\mathbf{r} - \mathbf{r}') = e/4 \pi \varepsilon_0 | \mathbf{r} - \mathbf{r}'|$ 
and exchange--correlation terms, with the last being discarded in this work.
Exchange terms are also ignored in the irreducible polarizability, leading to its non-interacting limit $\chiirr \to \chi^0$, where only the Hartree diagram is kept (leaving out the Fock terms). The random-phase approximation (RPA) counterpart of the full polarizability $\chi \to \chirpa$ can therefore be obtained by solving the Dyson equation~\cite{Bruus_Flensberg_2004,Martin_Reining_Ceperley_2016}
\begin{equation}
\label{eq:Dyson_eq_chi}
    \chirpa = \chi^0 + \chi^0 V_c \chirpa,
\end{equation}
where the Coulomb kernel $V_c$ includes only the bare Coulomb potential.
In momentum space, the previous equation is a matrix equation for each momentum $\mathbf{q} \in$ BZ, with the Coulomb kernel having matrix elements $(V_c)_{\mathbf{G} \mathbf{G}'} = v_c(\mathbf{q} + \mathbf{G}) \delta_{\mathbf{G} \mathbf{G}'}$, where the expression for~the Fourier transformed Coulomb potential depends on the number of dimensions.
Typically, \textit{ab initio} calculations \markup{exclude} exchange--correlation contributions to the Coulomb kernel~\cite{BerkeleyGW_2012,GPAW_review_2024,Martin_Reining_Ceperley_2016}, and eventually other terms accounting for relativistic effects such as spin can be included, but doing so is beyond the scope of this work.
Furthermore, \textit{ab initio} calculations are done in 3D (using off-plane periodic images for 2D materials), whence the commonly used momentum-resolved Coulomb potential in the literature is the 3D one, for which

\begin{equation}
    \label{eq:3D_Coulomb_potential}
    v^{\mathrm{3D}}_c (\mathbf{q}) = \frac{e}{\varepsilon_0 |\mathbf{q}|^2 \Omega_\mathrm{\scriptscriptstyle UC}},
\end{equation}

\noindent with $\Omega_\mathrm{\scriptscriptstyle UC}$ denoting the volume of the unit cell. \markup{In fact, for 2D materials it is a truncated version of this potential that is used, as we explain in the Methods~\ref{subsec:Chi&epsilon_Methods} section, after Eq.~\eqref{eq:macroscopic_epsilon_definition}}.
However, \markup{until the end of that section, we operate strictly within a 2D framework, as information about the system in the $z$ direction is either inaccessible or deliberately disregarded}. Consequently, we use the 2D Fourier-transformed potential.
This is commonly done and has shown to produce good results, both in analytical calculations for the dielectric function in atomically thin materials, such as the semimetal graphene~\cite{Hwang_2007,Sohier_Calandra_Mauri_2015}, and for excitons~\cite{GPAW_review_2024,Alex_Xatu_2024,Ninhos_2024,Henriques_Amorim_Ribeiro_Peres_2022,Henriques_Epstein_Peres_2022,Henriques_Peres_2020,Quintela_Henriques_Tenorio_Peres_2022,Thygesen_2017,Cudazzo_Tokatly_Rubio_2011} in 2D semiconductors and insulators. 
To the best of our knowledge, a fully numerical and 2D treatment of the dielectric function has not been done. This is the topic of the next subsection.

In the manner described above, the inverse RPA dielectric function can be obtained through the definition
\begin{equation}
\label{eq:inverse_dielectric_matrix_RPA}
    \varepsilon^{-1}_{\mathbf{G} \mathbf{G}'} (\mathbf{q}) = \delta_{\mathbf{G} \mathbf{G}'} + \chirpa_{\mathbf{G} \mathbf{G}'} (\mathbf{q}) v_c(\mathbf{q} + \mathbf{G}') \,.
\end{equation}
Alternatively, the same quantity can be obtained by computing the dielectric function matrix directly, which within RPA and symmetrized is given by Eq.~\eqref{eq:dielectric_matrix_Results}, and subsequently inverting it. In this work, we adopt the latter approach, though we still consider the former worth mentioning.

The 2D version of the implementation of the dielectric function has a striking advantage compared to the 3D counterpart when it comes to numerical instabilities, namely when computing the dielectric matrix around the origin $\mathbf{q} \sim 0$ and exactly at the origin, when $\mathbf{q} = \mathbf{0}$. 
For very small momentum transfer, the polarizability scales as $q^2$ for 2D semiconductors and insulators~\cite{BerkeleyGW_2012}, a behavior we have already verified in the Results~\ref{subsec:Results_Polarizability} section. As a result, $v_c(q)\chi^0_{\mathbf{0} \mathbf{0}}(q) \sim \mathcal{O}(q)$ as $q \to 0$, which ensures that no numerical instability arises at the origin for the head matrix element of the dielectric function, $\varepsilon_{\mathbf{0} \mathbf{0}}(\mathbf{0})$.
In this way, we can avoid having to use numerical schemes just to treat the problem of computing directly this matrix element, unlike other implementations in the literature, for instance the BerkeleyGW one from Deslippe \emph{et~al.}~\cite{BerkeleyGW_2012}.
For this particular matrix element, it does not matter whether one implements the symmetric dielectric function as per Eq.~\eqref{eq:dielectric_matrix_Results} or its non-symmetric version \markup{$\varepsilon_{\mathbf{G} \mathbf{G}'}(\mathbf{q}) = \delta_{\mathbf{G} \mathbf{G}'} - v_c(\mathbf{q} + \mathbf{G}')\chi^0_{\mathbf{G} \mathbf{G}'}(\mathbf{q})$}.
But as we will see, to deal with the wing elements $\varepsilon_{\mathbf{G} \mathbf{0}}(\mathbf{0})$ and $\varepsilon_{\mathbf{0} \mathbf{G}}(\mathbf{0})$ it is more convenient---both technically and computationally---to use the symmetrized form of the dielectric function~\cite{Baldereschi_Tosatti_1978, Rasmussen_Schmidt_Winther_Thygesen_2016, Shishkin_Kresse_2006}. Whence, we have chosen to implement the expression~\eqref{eq:dielectric_matrix_Results} instead of its non-symmetric version, \markup{since it bears no difference in the physics}.
Exactly at the origin, there is general consensus in the literature that some form of regularization must be applied to the dielectric matrix. Naturally, the screened potential also requires regularization, although we defer that discussion to the Methods~\ref{subsec:BSE_methods} section. The regularization is essential to prevent divergences or undefined values in the matrix elements of the inverse dielectric function---and consequently in the screened potential---ensuring that subsequent calculations, such as determining excitonic energies and wavefunctions, are not compromised by numerical errors.

Based on the discussion above, one can immediately conclude that the head matrix element satisfies $\varepsilon_{\mathbf{0} \mathbf{0}} (\mathbf{0}) =~1$. For the wing elements of the inverse dielectric matrix of both forms $\varepsilon^{-1}_{\mathbf{G} \mathbf{0}}(\mathbf{q})$ and $\varepsilon^{-1}_{\mathbf{0} \mathbf{G}}(\mathbf{q})$, if the symmetrized dielectric function is adopted, they approach zero when $q \to 0$ with $\sim\!\sqrt{q}$ (see section~S.2 of the SI). In practical implementations, it is therefore reasonable to set $\varepsilon^{-1}_{\mathbf{G} \mathbf{0}}(\mathbf{0}) = \varepsilon^{-1}_{\mathbf{0} \mathbf{G}}(\mathbf{0}) = 0$. Nevertheless, it remains important to understand the precise behavior of this limit in order to correctly regularize the screened potential---a topic addressed in the following section.

Our implementation allows for making a plausible estimate of the static macroscopic dielectric function of 2D materials with reasonable computational effort, through the commonly adopted definition in the literature~\cite{Latini_Olsen_Thygesen_2015}
\begin{equation}
\label{eq:macroscopic_epsilon_definition}
    \varepsilon_{\mathrm{M}} (\mathbf{q}) \equiv \frac{1}{\varepsilon^{-1}_{\mathbf{0} \mathbf{0}} (\mathbf{q})} \neq \varepsilon_{\mathbf{0} \mathbf{0}} (\mathbf{q}) \,,
\end{equation}
stressing that it does not coincide with the head matrix element of the dielectric function due to the local-field effects~\cite{Adler_1962,Wiser_1963}.
This is confirmed with the analysis of the inverse dielectric matrix in section~S.2 of the SI. We have actually verified numerically \markup{(see Fig.~S3 of the SI)} that ignoring the local-field effects overestimates the screening by far, unless the momentum transfer is very small, for which $\varepsilon_{\mathbf{0} \mathbf{0}} (\mathbf{q}) \approx 1/\varepsilon^{-1}_{\mathbf{0} \mathbf{0}} (\mathbf{q})$.
In the continuously translational invariant case, these effects are not present, and in that case, the microscopic and macroscopic dielectric functions coincide.
Notably, definition \eqref{eq:macroscopic_epsilon_definition} remains valid beyond the static regime. However, computing the dielectric function as a function of frequency lies beyond the scope and purpose of this manuscript.

It is more than pertinent to stress here a key point in our work. Due to the 2D nature of our calculations, the previously defined macroscopic dielectric function is actually the 2D effective macroscopic dielectric function. This is not true in general in \textit{ab initio} calculations, where the calculations are in 3D, and the results for 2D systems have to be converged against the amount of vacuum between adjacent repeated images, $L_{\perp}$.
In \textit{ab initio} calculations, the bare Coulomb potential used in practice for 2D materials is a truncated one and not 
Eq.~\eqref{eq:3D_Coulomb_potential}, \markup{serving the purpose of eliminating the Coulomb interaction between consecutive repeated system images~\cite{PhysRevB.73.205119,PhysRevB.73.233103}}.  \markup{Otherwise}, the calculations would always have to be converged with $L_{\perp}$. To argument in favor of our calculations, a truncation of the Coulomb potential in real space is unnecessary, as we work directly with its 2D Fourier transform. Furthermore, the macroscopic dielectric function defined by Eq.~\eqref{eq:macroscopic_epsilon_definition} is always one at zero momentum, and by consequence the head element of the inverse dielectric matrix is unity too. This is so by virtue of the polarizability reaching zero quadratically in momentum when the momentum approaches zero, but the Coulomb potential diverges with $1/q$. In contrast, in \textit{ab initio} calculations the macroscopic in-plane dielectric function at zero momentum is unity \markup{only if the truncated Coulomb potential is used~\cite{Huser_Olsen_Thygesen_2013}}. Moreover, the definition \eqref{eq:macroscopic_epsilon_definition} for 3D systems is valid only if we interpret that macroscopic dielectric function as a bulk one, while for 2D systems some sort of off-plane average has to be performed to obtain the true effective 2D macroscopic dielectric function~\cite{Latini_Olsen_Thygesen_2015,Diana_PhD_thesis,da_Jornada_PhD_thesis}.

Based on the analysis of our numerical macroscopic dielectric function in the Results~\ref{subsec:dielectric_function_results} section, we conclude that our strictly 2D computation of the microscopic dielectric function---followed by its inversion---naturally yields a macroscopic dielectric response that is inherently 2D. In contrast to 3D first-principles methods, our approach avoids off-plane image contributions entirely, eliminating the need for any off-plane averaging to define an effective 2D dielectric function. The macroscopic response emerges directly and unambiguously from our 2D framework. The trade-off is that with a strict 2D approach, one cannot correctly capture the dielectric function beyond the small-$q$ limit \markup{in quantitative terms}. To improve the accuracy of the dielectric properties of 2D systems using a 2D framework for the dielectric function, we can take the off-plane distances within the monolayer into account through a mixed representation for screening, mixing the momentum dependence of the dielectric function with the real-space dependence. \markup{Then, an off-plane average is necessary to obtain an effective 2D macroscopic dielectric function}. But fortunately, we can do so analytically. That is the subject of the next section.

\subsection{Quasi-2D approach for the dielectric function}
\label{subsec:Q2D_Methods}

In this part of the text we discuss the quasi-two-dimensional (Q2D) approach for the dielectric function adopted in this manuscript. It allows for incorporating the electronic structure pertaining to the off-plane physics. While the system is infinite (and periodic) along the $x$ and $y$ directions, which permits the electrons in Bloch states to propagate along the crystal, the electrons are confined along the $z$ direction.

Unlike in \textit{ab initio} approaches, where the system is periodic in all three Cartesian axis, herein the momentum is not a good quantum number for describing the delocalization of the electrons along the material thickness. Therefore, we are not allowed to perform a full 3D Fourier transform on the real-space (RPA) dielectric function given by

\begin{equation}
\label{eq:dielectric_function_real_space}
    \varepsilon (\mathbf{r},\mathbf{r}') = \delta(\mathbf{r} - \mathbf{r}') - \int  v_c(\mathbf{r} - \mathbf{x}) \chi^0 (\mathbf{x},\mathbf{r}') \, \mathrm{d} \mathbf{x} \,,
\end{equation}

\noindent where, for simplicity, we have omitted the dynamic nature of the dielectric function. Regardless of the dimensionality (that is, the number of dimensions along which the material is infinite) of the system, the real-space integral is over a volume region; the atomic orbitals of the individual atoms in the crystal have a 3D spatial structure, and the same can be said about the electron density.

In 3D systems, it is possible to work with a dielectric function defined in the 3D reciprocal space. The identity~\eqref{eq:dielectric_matrix_Results} still holds, but both the momentum $\mathbf{q}$ and reciprocal lattice vectors $\mathbf{G}$ and $\mathbf{G}'$ are 3D, having a $z$ component in reciprocal space. Also, the bare Coulomb potential follows Eq.~\eqref{eq:3D_Coulomb_potential}. This is the approach commonly found in the literature on computational methods for dielectric screening.

In our formalism, a 3D Fourier transform of Eq.~\eqref{eq:dielectric_function_real_space} is not useful. Instead, we perform a partial 2D Fourier transform of the real-space-resolved dielectric function, and keep track of the spatial $z$-dependence at the same time. Then, we work with the dielectric function in a mixed representation that reads

\begin{align}
    \label{eq:dielectric_function_mixed_representation}
    &\varepsilon_{\mathbf{G} \mathbf{G}'}(\mathbf{q},z,z') = \delta_{\mathbf{G} \mathbf{G}'} \delta(z-z') - \notag \\ 
    &\frac{e}{2\varepsilon_0 |\mathbf{q} +\mathbf{G}| \Omega_{\mathrm{UC}}} \int_{-\infty}^{\infty} \mathrm{d} z'' \mathrm{e}^{-|\mathbf{q} +\mathbf{G}||z-z''|} \chi^{0}_{\mathbf{G} \mathbf{G}'}(\mathbf{q},z'',z')\,,
\end{align}

\noindent where both (in-plane) momentum and $z$ variables are present, \markup{and we keep the asymmetric form for now}. The vectors $\mathbf{G}$ and $\mathbf{G}'$ are vectors of the reciprocal lattice, which is 2D, and the polarizability $ \chi^{0}_{\mathbf{G} \mathbf{G}'}(\mathbf{q},z,z')$ appears also in the mixed representation. To obtain the dielectric response, it is convoluted in the $z$ variable with the bare Coulomb potential in the mixed representation given by

\begin{equation}
    \label{eq:Coulomb_potential_mixed}
    v_{c}(\mathbf{q},z-z') =  \frac{e}{2 \varepsilon_0 q} \mathrm{e}^{-q |z-z'|}\,.
\end{equation}

The previous expression for the Coulomb potential is the main ingredient in our Q2D formalism. It is not hard to show that the previous expression is the Fourier transform of the position-resolved Coulomb potential along the $xy$ plane  in real space. In the limit $z=z'$, we recover the strict 2D Fourier transform of the Coulomb potential.

The polarizability in the mixed representation, which we will call the $(\mathbf{q},z)$-representation from now on, is defined as the 2D Fourier transform of the real-space-resolved polarizability $\chi^0(\mathbf{r},\mathbf{r}')$, where the in-plane spatial integrals are over the entire crystal (that is finite, and using periodic boundary conditions). To have a discussion as self-contained as possible, we provide the general expression for the non-interacting, irreducible RPA polarizability in real space. If we also want to include the frequency dependence of the polarizability, then we can write~\cite{Cohen_Louie_2016}

\begin{equation}
\label{eq:polarizability_real_space}
    \chi^0 (\mathbf{r},\mathbf{r}';\omega) = \sum_{i,j} (f_i - f_j) \frac{\phi_i^*(\mathbf{r}) \phi_j(\mathbf{r}) \phi_j^*(\mathbf{r}') \phi_i(\mathbf{r}')}{\epsilon_i - \epsilon_j + \hbar \omega + \mathrm{i} \eta}\,,
\end{equation}

\noindent where $\eta$ is a small infinitesimal parameter, $i$ and $j$ represent a set of quantum numbers, $\phi_i(\mathbf{r})$ is the $i$th state with eigenenergy $\epsilon_i$, and $f_i$ is the Fermi--Dirac distribution~\cite{Huang} computed at the energy $\epsilon_i$.
The states $\phi_i(\mathbf{r})$ are obtained within a mean-field approach, such as within HF~theory or the Kohn--Sham framework of DFT. In our work-flow, the band structure calculations are done prior to the screening calculations to obtain the model Hamiltonian for the material at hand. Then, we identify the states $\phi_i(\mathbf{r})$ with the Bloch states $\psi_{n \mathbf{k}}(\mathbf{r})$ in the LCAO approximation as per~\eqref{eq:Bloch_state_LCAO}, and the quasi-particle energies $\epsilon_i$ with $\epsilon_{n\mathbf{k}}$.

In this article, we consider zero absolute temperature (or a low enough temperature such that conduction states in a semiconductor remain unpopulated), and work in the static limit $\omega \to 0$ (which also allows to set $\eta = 0$). This simplifies our calculations. By performing an in-plane Fourier transform on~\eqref{eq:polarizability_real_space} in these approximations and applying the point-like orbital approximation, we obtain

\begin{widetext}
   \begin{align}
   \chi^{0}_{\mathbf{G} \mathbf{G}'}(\mathbf{q},z,z') = &\,\frac{1}{N} \sum_{v c,\mathbf{k} \sigma}\Bigg\{ \frac{1}{\epsilon_{v \mathbf{k}} - \epsilon_{c \mathbf{k}+\mathbf{q}}} \cdot \nonumber\\  
   &\sum_{i\alpha}\Big[C_{i\alpha}^{c \mathbf{k}+\mathbf{q}} \left(C_{i\alpha}^{v \mathbf{k}}\right)^{*} \mathrm{e}^{-\mathrm{i} (\mathbf{q} +\mathbf{G}) \vdot\mathbf{t}_{i}} \delta(z - t_{i,z}) \Big] \sum_{j\beta}\Big[C_{j\beta}^{v \mathbf{k}}\left(C_{j\beta}^{c \mathbf{k}+\mathbf{q}}\right)^{*} \mathrm{e}^{\mathrm{i} (\mathbf{q} -\mathbf{G}') \vdot \mathbf{t}_{j}} \delta(z' - t_{j,z}) \Big] - \nonumber\\
     &\frac{1}{\epsilon_{c \mathbf{k}} - \epsilon_{v \mathbf{k}+\mathbf{q}}} \cdot \nonumber \\
  &\sum_{i\alpha}\Big[\left(C_{i\alpha}^{c \mathbf{k}}\right)^{*}C_{i\alpha}^{v \mathbf{k}+\mathbf{q}}\, \mathrm{e}^{-\mathrm{i} (\mathbf{q} +\mathbf{G}) \vdot\mathbf{t}_{i}} \delta(z - t_{i,z}) \Big] \sum_{j\beta} \Big[\left(C_{j\beta}^{v \mathbf{k}+\mathbf{q}}\right)^{*}C_{j\beta}^{c \mathbf{k}}\, \mathrm{e}^{\mathrm{i} (\mathbf{q} +\mathbf{G}') \vdot \mathbf{t}_{j}} \delta(z' - t_{j,z})\Big] \Bigg\} \,.
\end{align} 
\end{widetext}

\noindent To obtain this result, one has to separate the $\mathbf{r}_\parallel$ and $z$ variables when doing the point-like orbital approximation as $\delta(\mathbf{r} - \bR - \mathbf{t}_i) = \delta(\mathbf{r}_\parallel - \bR - \mathbf{t}_{i,||}) \delta(z - t_{i,z})$, and in analogy we do the same for the primed variables $\mathbf{r}_\parallel'$ and $z'$. This expression for the polarizability in the $(\mathbf{q},z)$-representation is the one we use in Eq.~\eqref{eq:dielectric_function_mixed_representation} for the dielectric function. In this way, we are able to take into account the finite thickness of the material, that we will denote by $d_{\perp}$, and the out-of-plane position $t_{i,z}$ of the atoms composing the unit cell motif. By averaging the dielectric function over the monolayer thickness, we can define an effective Q2D microscopic dielectric function as

\begin{equation}
     \label{eq:quasi_2D_averaged_dielectric_function}
    \bar{\varepsilon}_{\mathbf{G} \mathbf{G}'}(\mathbf{q}) \equiv \frac{1}{d_{\perp}} \int_{-d_{\perp}/2}^{d_{\perp}/2} \int_{-d_{\perp}/2}^{d_{\perp}/2} \varepsilon_{\mathbf{G} \mathbf{G}'}(\mathbf{q},z,z')\,\mathrm{d}z \, \mathrm{d}z' ,
\end{equation}

\noindent where the integrals can be performed exactly by virtue of the properties of the Dirac delta-function. As dictated by the definition~\eqref{eq:macroscopic_epsilon_definition}, we can define an effective Q2D macroscopic dielectric function as $\varepsilon_{\mathrm{M}} (\mathbf{q}) = (\bar{\varepsilon}^{-1}_{\mathbf{0} \mathbf{0}}(\mathbf{q}))^{-1}$, where $\bar{\varepsilon}^{-1}_{\mathbf{0} \mathbf{0}}$ denotes the head element of the matrix inverse of~\eqref{eq:quasi_2D_averaged_dielectric_function}. Using this new definition for the macroscopic dielectric function was how we obtained the results in Fig.~\ref{fig:macroscopic_2D_epsilon}. \markup{In fact, the microscopic Q2D dielectric function we implemented is a symmetric version of Eq.~\eqref{eq:quasi_2D_averaged_dielectric_function}, just as we did for the 2D calculations. Its expression is given in Eq.~(S.25) in section~S5 of the SI, where one can see that the symmetric version of $\bar{\varepsilon}_{\mathbf{G} \mathbf{G}'}(\mathbf{q})$ enjoys the same symmetries as those of the strict 2D dielectric function}.

Solely by analytical calculations, we can show that the averaged Q2D dielectric function of Eq.~\eqref{eq:quasi_2D_averaged_dielectric_function} coincides with its strict 2D counterpart from Eq.~\eqref{eq:dielectric_matrix_Results} in  the zero thickness limit. Using math notation, we can write

\begin{equation}
    \label{eq:Q2D_dielectric_function_zero_thickness_limit}
    \varepsilon_{\mathbf{G} \mathbf{G}'}(\mathbf{q}) = \lim_{d_{\perp}\!\to 0} \bar{\varepsilon}_{\mathbf{G} \mathbf{G}'}(\mathbf{q})\,.
\end{equation}

\noindent This result allows us to define the long-wavelength or low-$q$ limit as the values of $q$ for which $q d_{\perp} \ll 1$. Furthermore, it explains why in this limit the 2D, Q2D and \textit{ab initio} macroscopic dielectric functions agree with each other. \markup{Both 2D and Q2D dielectric functions were used to compute the excitons as well, albeit within an approximate approach to build the Bethe--Salpeter Hamiltonian in the Q2D case}. More specifically, the matrix elements of the inverse dielectric function were used to compute the interaction matrix elements that enter in the BSE, the subject of the next section.

\subsection{Bethe--Salpeter equation}
\label{subsec:BSE_methods}

We assume that the excitonic state can be expressed as a linear combination of unbound electron--hole pair states, where the electron occupies a single-particle state in a conduction band $c$ and the hole corresponds to the absence of an electron in a valence band $v$. Formally, $|X\rangle = \sum_{\mathbf{k},v,c} A_{vc}(\mathbf{k}) c^\dagger_{c,\mathbf{k}} c_{v,\mathbf{k}} |\mathrm{FS}\rangle$, where the sum runs over pairs of valence and conduction bands, and only vertical transitions (same $\mathbf{k}$) are considered. Here, $|\mathrm{FS}\rangle$ denotes the ground state of the semiconductor or insulator (the Fermi sea), \markup{and $c^{\dagger}/c$ denotes the fermionic creation/annihilation operator}.
The exciton energy $E_X$ can then be determined by solving the BSE, which, within the Tamm--Dancoff approximation, reads~\cite{Grosso_Parravicini_2014}:
\begin{multline}
    \label{eq:Bethe_Salpeter_equation}
    \left(\epsilon_{c \mathbf{k}} - \epsilon_{v \mathbf{k}}\right) A_{vc}(\mathbf{k}) \\+ \sum_{c',v',\mathbf{k}'} K_{vc,v'c'}(\mathbf{k},\mathbf{k}')  A_{v'c'}(\mathbf{k}') = E_X A_{vc}(\mathbf{k}) \,.
\end{multline}
Here, the first term on the left-hand side accounts for the vertical transition between single-particle states (we assume these to be the KS eigenstates, where the quasi-particle corrections are already included). The second one accounts for the electron--hole interaction, is quantified by matrix elements of the many-body interaction kernel $K = -(D-X)$, that includes a direct term $D$ and an exchange term $X$. For instance, the direct term writes
\begin{multline}
\label{eq:D_matrix_element_definition}
    D_{vc,v'c'}(\mathbf{k},\mathbf{k}') \\= \int\mathrm{d}\mathbf{r}\, \mathrm{d}\mathbf{r}' \psi_{c \mathbf{k}}^*(\mathbf{r}) \psi^*_{v \mathbf{k}'}(\mathbf{r}') W(\mathbf{r},\mathbf{r}')  \psi_{c'\mathbf{k}'}(\mathbf{r}) \psi_{v'\mathbf{k}}(\mathbf{r}')
\end{multline}
in the approximation of static screening. The exchange term is obtained by simply exchanging $c',\mathbf{k}'$ with $v,\mathbf{k}$. Unlike the BSE derived from MBPT, where the exchange term is typically treated as unscreened, our approach does not require the exchange term to be unscreened, as justified in Ref.~\onlinecite{Alex_Xatu_2024}.

We model the static screened potential through the dielectric function in momentum space that we have computed previously. The relation between the screened Coulomb potential and the bare one is established through a Dyson equation, analogous to Eq.~\eqref{eq:Dyson_eq_chi}, that within RPA writes~\cite{Cohen_Louie_2016}
\begin{equation}
    W = V_c + V_c \chi^{0} W .
\end{equation}
Upon solving for $V_c$ in terms of $W$, defining $\varepsilon = \mathds{1} - V_c \chi^0$ and inverting the relation, in momentum space we can write
\begin{equation}
    \label{eq:screened_potential_def}
    W_{\mathbf{G} \mathbf{G}'}(\mathbf{q}) = \varepsilon^{-1}_{\mathbf{G} \mathbf{G}'}(\mathbf{q}) v_c(\mathbf{q} + \mathbf{G}') \,.
\end{equation}
This relation holds in both 3D and 2D. Within our strict 2D approach, $\varepsilon^{-1}_{\mathbf{G} \mathbf{G}'}(\mathbf{q})$ is the $(\mathbf{G} \mathbf{G}')$ matrix element of the inverse of the dielectric matrix as per \eqref{eq:dielectric_matrix_Results}, at the point $\mathbf{q}$ in the BZ. We implement the symmetrized dielectric matrix; therefore, the correct relation is $ W_{\mathbf{G} \mathbf{G}'}(\mathbf{q}) =\sqrt{v_c(\mathbf{q} + \mathbf{G})} \varepsilon^{-1}_{\mathbf{G} \mathbf{G}'}(\mathbf{q}) \sqrt{v_c(\mathbf{q} + \mathbf{G}')}$. Given the matrix elements of the screened potential and the expansion of the Bloch states in a basis of localized orbitals as in Eq.~\eqref{eq:Bloch_state_LCAO}, the direct matrix element in Eq.~\eqref{eq:D_matrix_element_definition}, within the point-like orbital approximation and after Fourier transforming and simplifying the screened potential, is given by
\begin{equation}
    \label{eq:direct_interaction_term}
     D_{vc,v'c'}(\mathbf{k},\mathbf{k}') = \frac{1}{N} \sum_{\mathbf{G} \mathbf{G}'} I^{\mathbf{G}}_{c \mathbf{k}, c' \mathbf{k}} W_{\mathbf{G} \mathbf{G}'}(\mathbf{k} - \mathbf{k}') (I^{\mathbf{G}'}_{v \mathbf{k}, v' \mathbf{k}'})^* \,,
\end{equation}

\noindent where each capital $I$ is a plane-wave matrix element between Bloch states, just as we defined initially in the Results~\ref{subsec:Results_Polarizability} section, and are evaluated by using Eq.~\eqref{eq:plane_wave_mel}.
The exchange term takes a similar form, but we do not include it in the calculations\mymarkup{, since we only consider the $\mathbf{Q}=\mathbf{0}$ case, where it only introduces a minor shift in the binding energies.} However, one must do so for non-vertical transitions where the exciton center-of-mass momentum $\mathbf{Q} \neq \mathbf{0}$ \mymarkup{or for more realistic predictions}.

\markup{It is worth noting that the definition in Eq.~\eqref{eq:screened_potential_def} is valid only in reciprocal space, which in our case is 2D. However, if we want to account for the non-vanishing material thickness along the lines of the Q2D approach from the previous section, then we need to somehow recover the $z$ variable. We do so approximately, as a rigorous handling of the off-plane degrees of freedom in calculating the excitons is outside the scope of this work. Exactly, the screened potential is a convolution in $z$ between the Coulomb potential $v_c(\mathbf{q}+\mathbf{G},z-z')$ and the inverse dielectric matrix in the mixed-representation $\varepsilon^{-1}_{\mathbf{G} \mathbf{G}'}(\mathbf{q},z,z')$ as}

\begin{equation}
    W_{\mathbf{G} \mathbf{G}'}(\mathbf{q},z-z') = \int \mathrm{d} z'' \varepsilon^{-1}_{\mathbf{G} \mathbf{G}'}(\mathbf{q},z,z'') v_c(\mathbf{q}+\mathbf{G}',z'',z') \,,
\end{equation}

\noindent \markup{if we work with a non-symmetric dielectric function. The symmetrization is left for the end after doing all approximations. We replace the inverse $(\mathbf{q},z)$-dielectric matrix by its off-plane average multiplied by $\delta(z-z')$, the former factors out of the convolution integral and the later renders the integral trivial. Then, we replace the average of the inverse dielectric function with the inverse of the averaged dielectric function as per Eq.~\eqref{eq:quasi_2D_averaged_dielectric_function}. In this way, the screened potential reads}

\begin{align}
    & W_{\mathbf{G} \mathbf{G}'}(\mathbf{q},z-z') \approx (\bar{\varepsilon}_{\mathbf{G} \mathbf{G}'}(\mathbf{q}))^{-1} v_c(\mathbf{q}+\mathbf{G}',z-z') \\ 
    &\approx \sqrt{\bar{v}_c(\mathbf{q}+\mathbf{G})} (\bar{\varepsilon}_{\mathbf{G} \mathbf{G}'}(\mathbf{q}))^{-1} \sqrt{\bar{v}_c(\mathbf{q}+\mathbf{G}')} \equiv \overline{W}_{\mathbf{G} \mathbf{G}'}(\mathbf{q}) \,, \notag
\end{align}

\noindent \markup{where we replaced the Coulomb potential in the mixed representation by its off-plane average we denote by $\bar{v}_c(\mathbf{q})$, and symmetrized the screened potential using the symmetrized version of the averaged Q2D dielectric function (as per Eq.~(S.25) of the SI).
In this way, we defined an effective Q2D screened potential $\overline{W}_{\mathbf{G} \mathbf{G}'}(\mathbf{q})$ that enjoys the same symmetries as those of its strict 2D counterpart. The expression for the averaged Coulomb potential $\bar{v}_c(\mathbf{q})$ can be found in Eq.~(S.26) at the end of section S.5 of the SI. Then, the interaction matrix elements have the same form as in the 2D approach, but with the 2D screened potential $ W_{\mathbf{G} \mathbf{G}'}(\mathbf{q})$ replaced by the Q2D effective one $\overline{W}_{\mathbf{G} \mathbf{G}'}(\mathbf{q})$ we have just defined. Using this method, we obtain the values for the exciton binding energies inside curved brackets in the diagonal cells of Table~\ref{tab:Exciton_binding_energy}.}

It has previously been demonstrated that evaluating the screened potential in real space leads to significantly faster convergence of the calculations compared to evaluating it in momentum space~\cite{Alex_Xatu_2024}. Conversely, it has also been reported that explicitly including the $\mathbf{q} = \mathbf{0}$ term improves the numerical convergence with respect to the number of $\mathbf{k}$-points used~\cite{Rasmussen_Schmidt_Winther_Thygesen_2016}. Here, we adopt a regularization scheme for the term $W_{\mathbf{0} \mathbf{0}} (\mathbf{q} = \mathbf{0})$ proposed in Ref.~\onlinecite{Huser_Olsen_Thygesen_2013}, which we discuss in detail in Sec.~S.2 of the SI. To our knowledge, this approach has not been applied previously.

For the sake of making the main text as self-contained as possible, we explain how we regularized the potential for the singular terms when $\mathbf{q} = \mathbf{0}$. As proposed in Ref.~\onlinecite{Huser_Olsen_Thygesen_2013}, the head matrix element of the screened potential at the origin is replaced by an average, integrating the screened potential in a sufficiently small region $\Omega_\Gamma$ centered at the origin $\Gamma$ of the BZ, that we take here as a small circle of radius $q_0$. The average reads
\begin{equation}
\label{eq:head_term_regularization}
    W_{\mathbf{0} \mathbf{0}} (\mathbf{0}) = \frac{1}{\Omega_\Gamma} \int_{\Omega_\Gamma} \mathrm{d} \mathbf{q} \, v_c(\mathbf{q}) \big[1 + \mathbf{q} \cdot \mathbf{\nabla}_{\mathbf{q}} \varepsilon^{-1}_{\mathbf{0} \mathbf{0}} (\mathbf{q})|_{\mathbf{q} = \mathbf{0}} \big] \,,
\end{equation}
and can be computed analytically by numerically evaluating the gradient appearing in the expression. We know that the head matrix element of the inverse dielectric function is linear in $\mathbf{q}$ for $\mathbf{q} \sim 0$, so that $\mathbf{\nabla}_{\mathbf{q}} \varepsilon^{-1}_{\mathbf{0} \mathbf{0}} (\mathbf{q})|_{\mathbf{q} = \mathbf{0}} = (-r^x_0,-r^y_0)$, where we use two screening parameters, one along the $x$ direction and the other along the $y$ direction (directions in reciprocal space, that is). Henceforth, $r^x_0$ and $r^y_0$ are estimated numerically to evaluate~\eqref{eq:head_term_regularization}. If the material is isotropic, which is the case of both \ch{hBN} and \ch{MoS2}, then $r^x_0 = r^y_0$, and only $r^x_0$ needs to be computed. 

The other matrix elements that require treatment by hand are the wing ones. However, we argue in section~S.2 of the SI that these matrix elements can be simply ruled out and take them as null: $W_{\mathbf{G} \mathbf{0}} (\mathbf{0}) = W_{\mathbf{0} \mathbf{G}} (\mathbf{0}) = 0$. This is a byproduct, and another advantage, of the Coulomb potential in our implementation diverging with $1/q$ instead of $1/q^2$. In this way, we avoid the need for performing numerical angular averages for these matrix elements as it is done in previous works~\cite{Rasmussen_Schmidt_Winther_Thygesen_2016}. Even though the low $q$-dependence, just as in Ref.~\onlinecite{Rasmussen_Schmidt_Winther_Thygesen_2016}, goes with the direction $\hat{\mathbf{q}} = \mathbf{q}/q$ along which we take the limit, in our case it does so linearly (see equations (S.22) and (S.23) of the SI). Hence, performing a numerical average (over a centrosymmetric region to fully account for eventual anisotropies) would yield a null result.

\section*{Acknowledgments}

We thank Nuno M. R. Peres for stimulating and insightful discussions. We thank Guilherme Janone for providing the Wannier model for \ch{MoS2} and M. Quintela for testing the beta version of XATU.
P.~N. acknowledges support by the Otto M{\o}nsted Foundation (grant No. 24-70-1631), which funded their research stay abroad at Universidad Autónoma de Madrid, and the Independent Research Fund Denmark (grant No. 2032-00045A), which funded their PhD studies.
The Center for Polariton-driven Light--Matter Interactions (POLIMA) is funded by the Danish National Research Foundation (project No.~DNRF165).
J.~J.~P. acknowledges financial support from MICIU/AEI/10.13039/501100011033/UE through Grants  PCI2026-177428-1 (M.ERA-NET program), PID2022-141712NB-C21, the “María de Maeztu” Programme for Units of Excellence in R\&D (CEX2023-001316-M), the Comunidad de Madrid within the Recovery, Transformation and Resilience Plan, and by NextGenerationEU programme from the European Union through the project “Disruptive 2D materials” (MAD2D-CM-UAM7), the Generalitat Valenciana through the Program Prometeo (2021/017), and the Naturgy Foundation. We also acknowledge computer resources and assistance provided by Centro de Computación Científica de la Universidad Autónoma de Madrid and RES resources (FI-2025-3-36, FI-2025-2-34, FI-2025-1-12, FI-2024-3-0010, FI-2024-2-0016, FI-2024-1-0038). 

\section*{Author Contributions}

P.~N. designed and conducted the research; A.~J.~U.-A. developed the first version of the code, introduced the first author to it and gave advice on the implementation of the new functionalities;  J.~J.~P. analyzed the data; C.~T., N.~A.~M. and J.~J.~P. supervised the research; and P.~N., A.~J.~U.-A., C.~T., N.~A.~M. and J.~J.~P. wrote the manuscript.

\section*{Data Availability}

The datasets generated and/or analyzed during the current study are not publicly available at this moment due to the merge with the official repository being work in progress, but are available from the corresponding author on reasonable request.
All the data is reproducible with the new version of the XATU code, which will be released soon, upon using the same convergence parameters one can find in the main text and SI.

\section*{Code Availability}

The code used to conduct our studies was developed on top of the first version of the XATU code, available at \url{https://github.com/xatu-code}.
The merge with the official repository is work in progress, and the new version will be released soon.

\section*{Competing Interests}

The authors declare no competing financial or non-financial interests.

\bibliography{references}

\end{document}